%% file: branch_pt_paper.tex
\documentclass[11pt]{article}
\usepackage{fancyhdr}
\usepackage{isomath}
\usepackage{amsmath}
\usepackage{amsbsy}
\usepackage{amssymb}
\usepackage{amscd}
\usepackage{amsfonts}
\usepackage{graphicx,color}
\usepackage{verbatim}
\usepackage{euscript}
\usepackage{alltt}
\usepackage{stmaryrd}
\usepackage{subcaption}
\usepackage{bigints}
\usepackage{url}

\input{temp-definitions}

\usepackage[margin=1in]{geometry}

\newenvironment{rcases}
  {\left.\begin{aligned}}
  {\end{aligned}\right\rbrace}

\date{}
\begin{document}
\title{
Continuum mechanics of moving defects in 
growing bodies}
\author{Amit Acharya\thanks{Department of Civil \& Environmental Engineering, and Center for Nonlinear Analysis, Carnegie Mellon University, Pittsburgh, PA 15213, email: acharyaamit@cmu.edu.} \and Shankar C. Venkataramani\thanks{Department of Mathematics, University of Arizona, Tucson, AZ 85721, email: shankar@math.arizona.edu.}
}
\maketitle

\begin{abstract}
\noindent Growth processes in many living organisms create thin, soft materials with an intrinsically hyperbolic geometry. These objects support novel types of mesoscopic defects -- discontinuity lines for the second derivative and {\em branch points} -- terminating defects for these line discontinuities. These higher-order defects move ``easily", and thus confer a great degree of flexibility to thin hyperbolic elastic sheets.  We develop a general, higher-order, continuum mechanical framework from which we can derive the dynamics of higher order defects in a thermodynamically consistent manner. We illustrate our framework by obtaining the explicit equations for the dynamics of branch points in an elastic body.
\end{abstract}

\section{Introduction}

\begin{figure}[htbp]
\centering
	\begin{subfigure}[t]{0.3\textwidth}
		\centering
		\includegraphics[height=1.8in]{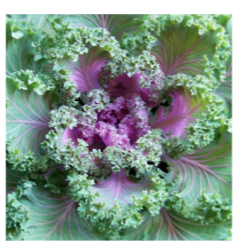}
		\caption{Kale ({\em Brassica Oleracea}).}
	\end{subfigure}
	~
	\begin{subfigure}[t]{0.3\textwidth}
		\centering
		\includegraphics[height=1.7in]{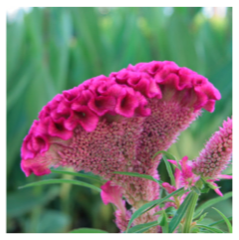}
		\caption{Cokscomb ({\em Celosia Cristata}) }
	\end{subfigure}
	~
	\begin{subfigure}[t]{0.3\textwidth}
		\centering
		\includegraphics[height=1.7in]{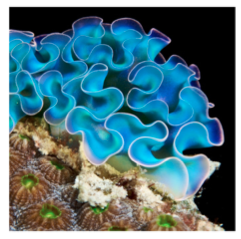}
		\caption{Sea-slug ({\em Elysia Cristata})}
	\end{subfigure}
	 \caption{Examples of naturally occurring non-Euclidean elastic sheets.}
	\label{fig:splashy}
\end{figure}

Hyperbolic sheets abound in nature (see Fig.~\ref{fig:splashy}). As Margaret Wertheim writes in her delightful essay ``Corals, crochet and the cosmos: how hyperbolic geometry pervades the universe"  \cite{wertheim2016corals} -- {\em We have built a world of largely straight lines -- the houses we live in, the skyscrapers we work in and the streets we drive on our daily commutes. Yet outside our boxes, nature teems with frilly, crenellated forms, from the fluted surfaces of lettuces and fungi to the frilled skirts of sea slugs and the gorgeous undulations of corals.} 

A natural question is  --  Why these shapes? 
One suggestion is that cells in living organisms proliferate to ``maximize" their number (area) subject to any applicable constraints \cite{wertheim2016corals} and this naturally results in hyperbolic geometries. 
This is a ``static" argument the relates the mechanisms of growth to the resulting (quasi-2D) {intrinsic geometry} of living organisms. In this paper, we attempt to go beyond this ``static" argument and develop models, based on thermodynamic considerations, to gain a quantitative understanding of the interplay between growth, mechanics and dynamics in soft objects. These models have the potential to describe the dynamical processes that result in the observed intricate three-dimensional (i.e. extrinsic) morphologies in nature. 

\begin{figure}[ht!]
   \centering
   \begin{subfigure}[t]{0.23\textwidth}
     \centering
     \includegraphics[trim={1.5cm, 1.5cm, 1.5cm, 1.5cm}, clip, width=\textwidth]{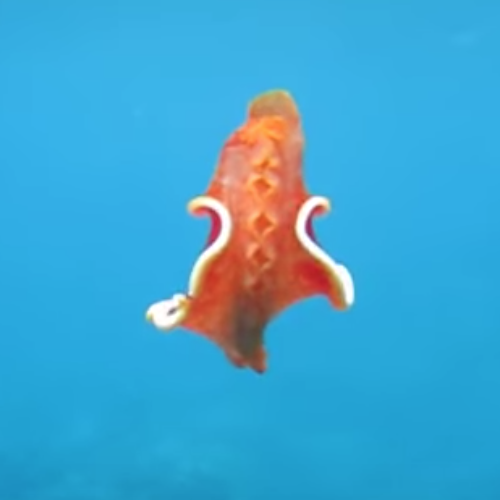}
   \end{subfigure}
   ~
   \begin{subfigure}[t]{0.23\textwidth}
     \centering
     \includegraphics[trim={1.5cm, 1.5cm, 1.5cm, 1.5cm}, clip, width=\textwidth]{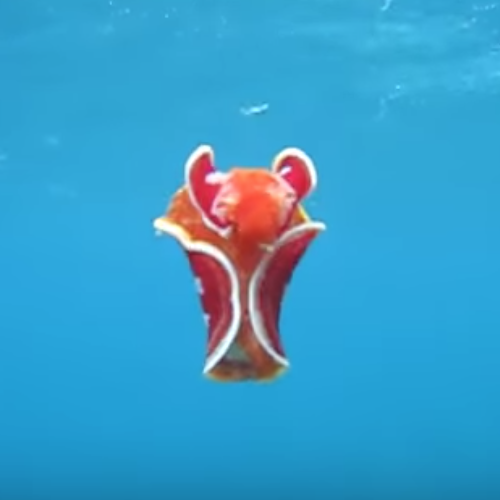}   \end{subfigure}
   ~
   \begin{subfigure}[t]{0.23\textwidth}
     \centering
     \includegraphics[trim={1.5cm, 1.7cm, 1.5cm, 1.3cm}, clip, width=\textwidth]{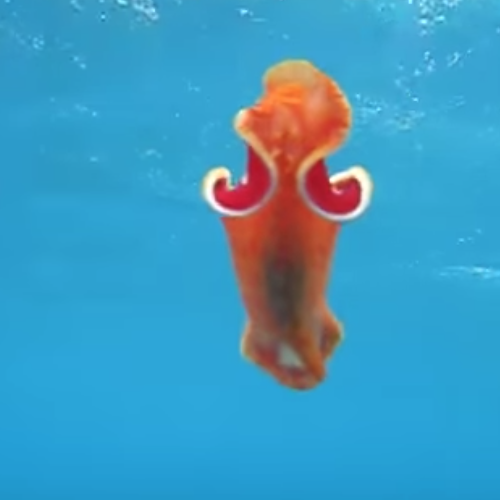}
   \end{subfigure}
   ~
   \begin{subfigure}[t]{0.23\textwidth}
     \centering
     \includegraphics[trim={1.5cm, 1.5cm, 1.5cm, 1.5cm}, clip, width=\textwidth]{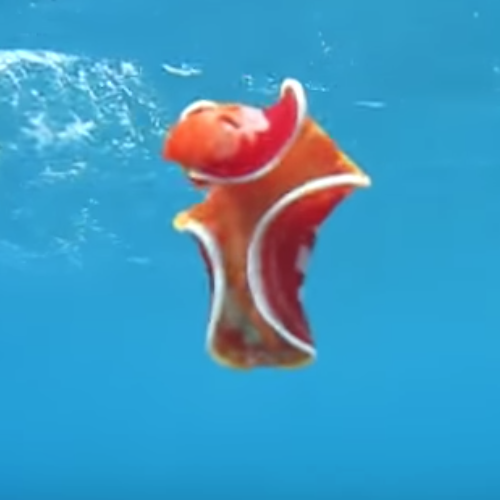}
   \end{subfigure} 
 \caption{A free swimming sea slug {\em Hexabranchus Sanguineus}. The frames are 2s apart. Images used with permission from the copyright holders of the original video~\protect{\cite{wavelength}}.}
   \label{fig:nudibranch}
\end{figure}

Particularly striking examples dynamical behaviors in organisms with differential growth (hyperbolic geometries) occur in sea slugs ({\em Nudibranchia})  and marine flatworms ({\em Polycladida}). These marine invertebrates are found in many environments, particularly in coral reefs. While most of them crawl on the sea floor,  a few are capable of free swimming \cite{newman2003marine}. They move/swim by sending waves of undulations from the front to the back along their skirts (for sea slugs) or across their entire body (for flatworms). 
Fig.~\ref{fig:nudibranch} shows 4 frames from a video of a free-swimming sea slug \cite{wavelength}. 
The geometry of the slug is clearly hyperbolic. It has multiple undulations 
and undergoes significant bending deformations in the course of one swim cycle. 
While it is hard to quantify the strains within the organism it is not unreasonable to consider them small in comparison to the 
obvious large 
rotations/twist of the body.

In a different context, the interplay between growth and dynamics is also relevant to the development of leaves, flowers and other plant tissues that can be modeled as thin laminae \cite{LiangMaha2009,boudaoud2010plants,LiangMaha2011,goriely2017morphogenesis,sharon2018mechanics}. Laboratory experiments using hydrogels \cite{KES07,halftone-gels} have led to a semi-quantitative understanding of time-dependent, dissipative deformations of thin soft materials with a prescribed prestrain. In living organisms, however, the prestrain is not prescribed {\em a priori}, and how the prestrain development may be related to mechanics is not clear. Complicated physico-chemical processes are involved that need to be incorporated into mathematical models. It therefore seems reasonable to derive systematic constraints on the mathematical description based on a careful consideration of the non-standard kinematics involved and the general principles of continuum thermomechanics.

Earlier work, reviewed briefly in Sec.~\ref{sec:statics}, implicates higher-order defects, in contrast to disclinations and dislocations, as playing a key role in the mechanics of intrinsically hyperbolic elastic sheets \cite{GV2011,GV2013,EPL_2016,toby_todo}. This  points to the need for tools to  describe the evolution of (terminating) discontinuities of the second-gradient of the displacement field - when viewed at the macroscopic scale - for a proper description of the soft material deformations involved. It turns out that, within a continuum mechanical perspective, this fits in nicely within the question of describing the coupled mechanics of discontinuities and singularities of the elastic displacement field and its higher derivatives up to order three. This is the question that is addressed in this paper.


While, as evident from Fig.~\ref{fig:nudibranch},  it is natural, and necessary, to consider unrestricted finite deformations when dealing with soft materials, we restrict attention to `small deformation' kinematics in this first effort due to the extra subtleties involved with higher order defect kinematics. Thus, we consider deformations of a fixed reference configuration that may or may not be stress-free. When the configurations attained by the system remain in close proximity to this fixed configuration, this is an adequate assumption. We consistently invoke Occam's razor as a guiding principle in our development - for instance, we restrict to the use of only ordinary stresses and couple stresses since forces and moments are the only agents of mechanical stimuli that we have some intuition for. Similarly, if branch point and surface defect velocities are to be the only dissipative mechanisms requiring constitutive specification without involving their spatial derivatives, then it turns out that the appropriate variables for the analysis of thermodynamics is in terms of the `singular parts' of the first and higher order displacement gradients, instead of the more natural singular parts of the corresponding elastic distortion gradients that naturally arise in the analysis of defect kinematics. This is in sharp contrast to dislocation and g.disclination mechanics \cite{acharya2015continuum} where this distinction does not arise because of the relatively lower order kinematics involved. We develop the relationship between the two types of entities in this paper.


\section{Statics and equilibria of non-Euclidean elastic sheets}  \label{sec:statics}

One approach to modeling the mechanics of a growing hyperelastic body, borrowed from the literature of finite elastoplasticity, is to assume a reference configuration $\mathcal{S}$ and a deformation $y:\mathcal{S} \to \mathbb{R}^3$ along with a multiplicative decomposition of the deformation gradient $F = \nabla y$ as $F = E G$ (or $F = F^e F^p$ in the plasticity literature) where the two-point tensor $G$ models the effect of the growth processes in the material and $E$ is the ``residual" elastic deformation \cite{goriely2017morphogenesis}. The energy of the configuration defined by $y$ is then given by $\int W(E) = \int W(F G^{-1})$, where $W$ denotes a hyperelastic energy density, vanishing on $SO(3)$ \cite{goriely2017morphogenesis,lewicka2014models}. In particular, the material is ``stress-free" if $FG^{-1} \in SO(3)$, although, for a general $G$, there might not be any deformation of the body $y:\mathcal{S} \to \mathbb{R}^3$ whose gradient is a rotation times $G$. Such objects, with no stress-free configurations in $\mathbb{R}^3$, lead to {\em incompatible elasticity.}

The non-Euclidean formalism of thin sheet elasticity  \cite{efrati2009elastic} is a reduced dimensional description of thin elastically incompatible objects. The reference manifold $\mathcal{S} = \Omega \times [-\frac{t}{2},\frac{t}{2}]$, where $t$, the thickness, is ``small" compared to the ``in-plane'' dimensions of the center surface $\Omega \subset \mathbb{R}^2$. In this setting, the effect of the growth has a reduced dimensional description as a 
2-manifold $(\Omega,g,b)$ where $g,b$ are symmetric $(0,2)$ tensors. These tensors denote, respectively,  the  
 `target' 1st and 2nd fundamental forms of the stress-free state of the sheet, pulled back to the reference manifold \cite{efrati2009elastic}.  
%
This framework also applies to 
{\em incompatible elasticity} \cite{ben-amar2005growth,lewicka2014models,bhattacharya2016plates} 
where, in general, there exists no deformation $f: \Omega \to \mathbb{R}^3$ realizing a surface in ambient three-dimensional space whose first and second fundamental forms match (the push-forward of) the targets $g,b$ (by $f$), 
i.e., incompatible sheets have no stress-free configurations in our three dimensional space. 

Assuming the Kirchhoff-Love hypothesis \cite{solid-mech-book}, 
so that the (3D) deformation of a thin sheet is determined by the (2D) mapping $y:\Omega \to \mathbb{R}^3$ on the center surface. 
This allows for and asymptotic expansion of the elastic energy as a sum of stretching and bending contributions \cite{efrati2009elastic,efi} :
\begin{align}
E^t[y] 
=\int_{\Omega} \left[t \,Q_3(\nabla y^T\cdot \nabla y-g)+\frac{t^3}{12} \,Q_3(\nabla y^T\cdot \nabla N -b)\right]\,dA,
\label{eq:elastic}
\end{align}
where the oriented normal field $N: \Omega \to S^2$, also called the {\em Gauss Normal map} \cite{stoker},  is obtained from $\nabla y^T \cdot N = 0$. $Q_3$ is a non-degenerate quadratic form, on symmetric $2 \times 2$ matrices, that depends on the Poisson's ratio $\nu$ of the material \cite{efrati2009elastic},
and $dA$ is the area element on $(\Omega,g)$.

For various choices of $g$ and $b$ and boundary conditions, the energy functional~\eqref{eq:elastic} describes a variety of phenomena in thin sheets, 
including multiple-scale buckling in free sheets with `excess length' near an edge,~e.g.~torn~plastic~or~flat leaves treated with an Auxin near the edge \cite{eran,eran2,eran2004leaves}.
The excess length near the edge 
is modeled by a metric $g$ with negative intrinsic curvature \cite{efi}. 

Starting with a fully 3D elastic energy, Lewicka and Pakzad \cite{lewicka2011scaling} have obtained a reduced dimensional model for the limit $t \to 0$ using $\Gamma$--convergence. In particular, they showed  that 
$$
\underset{t \to 0}{\Gamma \mhyphen \lim} \, \frac{E^t[y]}{t^3} = \mathcal{E}^*[y] = \frac{1}{12} \begin{cases} \int_\Omega Q_2(\nabla y^T\cdot \nabla N -b)\, dA &   \mbox{if } \nabla y^T\cdot \nabla y-g = 0 \mbox{ a.e } \\ + \infty & \mbox{ otherwise.} \end{cases}
$$
for an appropriate quadratic form $Q_2$. This energy has clear similarities with the energy in~\eqref{eq:elastic}, although the details are somewhat different. 
Nonetheless, in either framework, the elastic energy scales like $t^3$ in the thin limit $t \to 0$ {\em if and only if}  there exist finite bending energy (mathematically $y \in W^{2,2}$) isometric immersions $y:(\Omega,g) \to \mathbb{R}^3$. 
To illustrate the physical import of this theorem, we remark that $W^{2,2}$ surfaces 
necessarily have a continuous tangent plane and normal  \cite{evans2} so they cannot contain 
sharp creases (folds), cone points (disclinations) or dislocations. Indeed the energy of elastic ridges \cite{lobkovsky,MR2023444,conti2008confining}, $E^t \sim t^{8/3}$,  and $d$-cones \cite{MbAYP97,Cerda-1999,olbermann2016d-cone}, $E^t \sim t^3 \log(1/t)$,  diverge on the scale $t^3$, 
although the limiting shapes are `asymptotic' isometries \cite{vella2015indentation,davidovitch2019geometrically} and arguably unstretched. 

This theorem is central to the work in this paper. We are mainly interested in the mechanics of thin objects whose energy scale is $O(t^3)$ as $t \to 0$ (which includes smaller energies scaling with powers of $t$ greater than 3). Such objects are naturally ``floppy" since they are governed by ``bending" and weaker forces, 
and the relevant defects are higher order in contrast to dislocations or (g.)disclinations.

\subsection{Branch points and lines of inflection} \label{sec:branch-points}

The preceding remark highlights the role of the {\em regularity} of isometries. Beyond the existence/non-existence of isometries, it is crucial whether a candidate isometry is in $W^{2,2}$. This motivates the problem:
\begin{equation}
\mbox{Find } y: \Omega \to \mathbb{R}^3  \mbox{ such that } \begin{cases} \nabla y^T\cdot \nabla y=g \text{ and } & \\ \mathcal{B} =  \int_\Omega Q(\nabla y^T\cdot \nabla N -b)\, dA < \infty, &
\end{cases}
\label{W22isometry}
\end{equation}
We have rigorous results showing that the problem~\eqref{W22isometry} is flexible and solutions are plentiful~\cite{EPL_2016,toby_todo} (with prescribed zero-traction and moment boundary conditions, i.e. for free sheets).  The proof is constructive, and uses ideas from Discrete Differential Geometry DDG \cite{ddg1,EPL_2016}. This lack of uniqueness in admissible static configurations with prescribed boundary conditions underscores the necessity of a dynamical model to `choose' between acceptable configurations and/or describe the transitions between multiple admissible states \cite{GV2013}.

If $y: \Omega \to \mathbb{R}^3$ is $C^1$, the Gauss Normal map is given by $\displaystyle{N = \frac{\partial_1 y \times \partial_2 y}{\|\partial_1 y \times \partial_2 y\|}}$, where $\displaystyle{\partial_{i} = \frac{\partial}{\partial x^i}}$  for (arbitrary) coordinates $(x^1,x^2)$ on $\Omega$. Further, if $y$ and $g$ are $C^2$, Gauss' {\em Theorema Egregium} implies that
~\eqref{W22isometry} is equivalent to the Monge-Ampere Exterior differential system (EDS) \cite{EDS,ivey2003cartan}:
\begin{equation}
N \cdot dy = 0, \qquad N^*(d\Omega) = \kappa \,dA, \qquad \kappa \equiv \kappa[g] \mbox{ is determined by } g,
\label{thmegrg}
\end{equation}
where $d \Omega$ is the area form on the sphere $S^2$ and $\kappa$ is the Gauss curvature.

\begin{figure}[htbp]
	\centering
	\begin{subfigure}[t]{0.55\textwidth}
		\centering
		\includegraphics[height=1.5in]{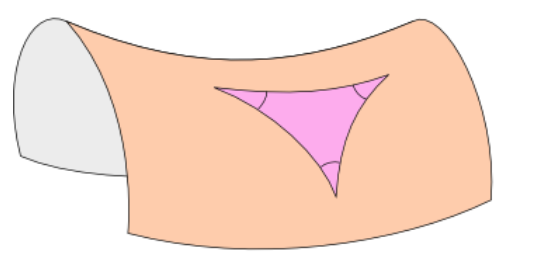}
		\caption{Smooth saddle surface}
		\label{fig:saddle}   
	\end{subfigure}
	~
	\begin{subfigure}[t]{0.4\textwidth}
		\centering
		\includegraphics[height=2in]{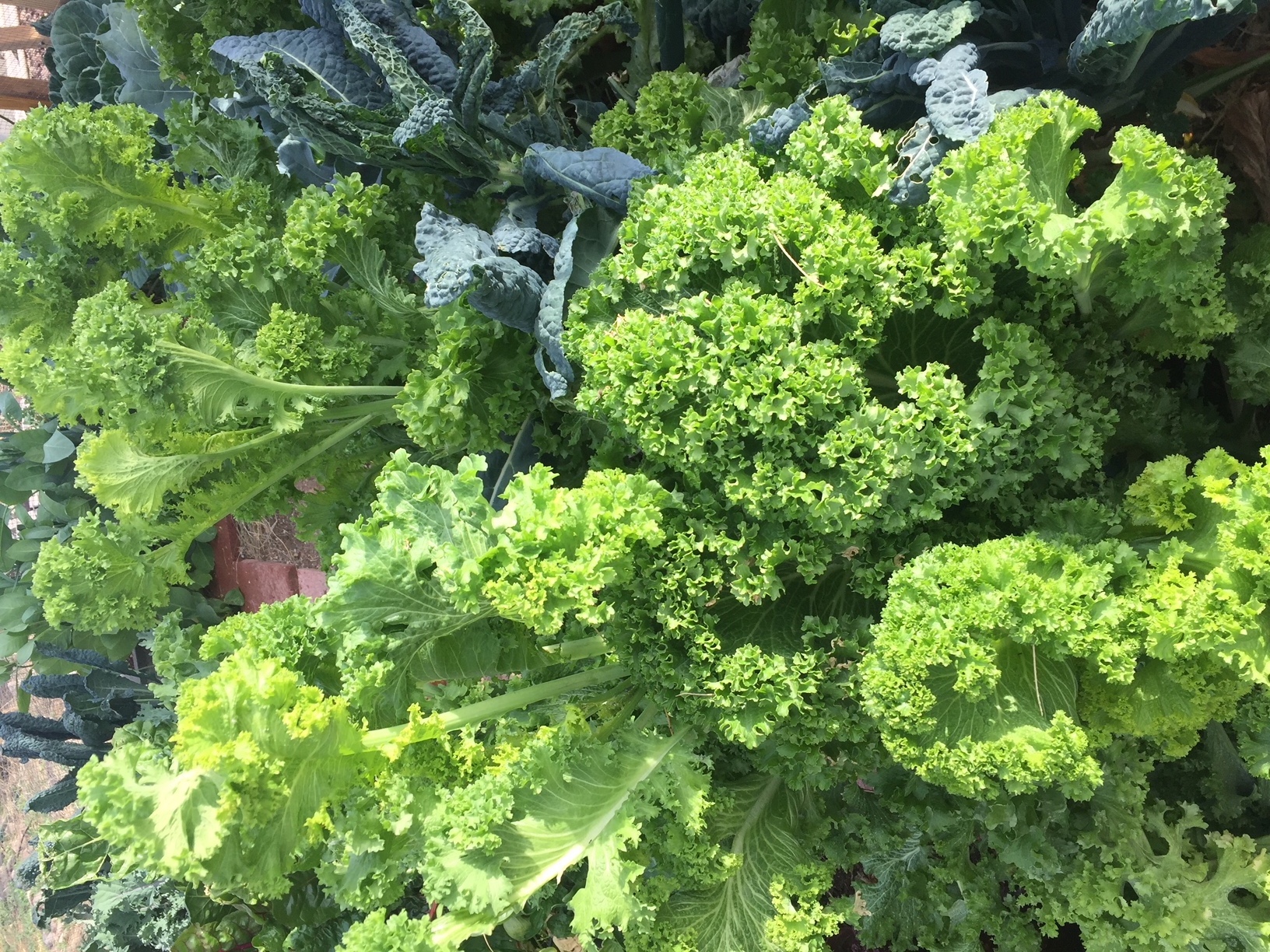}
		\caption{Curly Mustard. Image courtesy Joe Watkins.}
		\label{fig:mustard}
	\end{subfigure}
	 \caption{Hyperbolic surfaces in $\mathbb{R}^3$. The inscribed (geodesic) triangle in the smooth saddle has angles that sum up to less than $\pi$, illustrating the connection
between the extrinsic and intrinsic geometries --  Gauss' Theorema Egregium.}	
	 \label{fig:surfaces}
\end{figure}
Classical results in differential geometry imply that smooth solutions of~\eqref{thmegrg} with $\kappa < 0$ are hyperbolic surfaces and locally saddle shaped.
In contrast, the curly mustard leaf in Fig.~\ref{fig:mustard} is ``frilly", i.e buckled on multiple scales with a wavelength that refines (``sub-wrinkles") near the edge \cite{eran2004leaves}.
This ``looks" very unlike the smooth saddle 
 in Fig.~\ref{fig:saddle}. 
%
Why do we see frilly shapes in natural surfaces, as in Fig.~\ref{fig:mustard}, rather than the smooth saddles of Fig.~\ref{fig:saddle}? Indeed, any finite piece of a smooth hyperbolic surface can always be smoothly and isometrically embedded in $\mathbb{R}^3$  \cite{han-hong-book} as ``non-frilly" surfaces.

We have addressed this puzzle in recent work \cite{GV2011,GV2012,GV2013,EPL_2016,toby_todo}  and the short answer is that, for a given metric $g$, the frilly surfaces, somewhat counterintuitively, can have {\em smaller} bending energy than the smooth saddle. It is true that $C^2$ (twice continuously differentiable) hyperbolic surfaces are saddle-like near every point. We find a topological invariant \cite{toby_todo}, the index of a branch point - intimately related to the quantity $\int_\Sigma \widehat{\alpha}^{(3)} n\, da$ that emerges in Sec.\! \ref{sec:kin} and the quantity $\Gamma$ of Sec.\! \ref{sec:wein} - that distinguishes sub-wrinkled surfaces from saddles locally. With branch points, the surfaces are only $C^{1,1}$, but gain the additional flexibility to 
refine their buckling pattern, while lowering their energy \cite{EPL_2016}. This flexibility {\em is not available} to smooth saddles, and constitutes a key property of {\em branched} (sub-wrinkled) surfaces \cite{EPL_2016,toby_todo}. 

\begin{figure*}[htp]
\begin{center}
\includegraphics[width = 0.9 \linewidth]{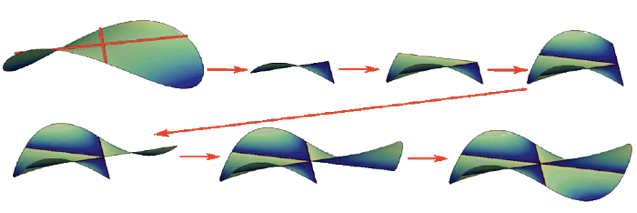}
\end{center}
\caption[Monkey-saddle]{\label{fig:monkey-saddle} A piecewise quadratic monkey saddle. We are grateful to John Gemmer for producing this figure for us.}
\end{figure*}

Figure~\ref{fig:monkey-saddle} shows a non-$C^2$ monkey saddle -- a piecewise quadratic surface made from 6 sectors congruent to the wedge 
$w = x^2 - 3 y^2, x \geq \sqrt{3} |y|$, patched together by odd reflections about the lines $x = \pm \sqrt{3} y, x = 0$ \cite{GV2011}.  
Although this surface is not $C^2$, it is indeed $W^{2,2}$ and 
has a continuous normal vector and bounded curvature everywhere. Its ``defects" include the point in the middle -- a {\em branch point} and the 6 rays through this point -- {\em lines of inflection}, which together constitute the {\em asymptotic skeleton} of the surface \cite{toby_todo}. This construction can be extended to generate $C^{1,1}$ hyperbolic surfaces with multiple distinct branch points, and an interesting question is how these defects interact with and influence each other \cite{EPL_2016}. 

Defects are of course ubiquitous in condensed matter systems. A key feature of defects in systems driven by a free energy is that the energy density typically diverges  in the vicinity of a ``bare" defect (and in some cases even the total energy diverges), and as a consequence, defects are always regularized, i.e. ``cored" in physical systems. This is true for dislocations and disclinations in elastic objects, for creases in crumpled sheets, for defects in liquid crystals and many other types of defects. Uniquely, branch points and lines of inflection do not carry a singular energy density 
\cite{GV2011,EPL_2016}, and thus do not ``need" a core for energetic reasons. %
Nonetheless, force and moment balance implies that these defects are indeed regularized into boundary layers, of width $t^{1/3}$, mediating jumps in the normal curvature \cite{GV2012}. 

Like other defects in condensed matter, branch points and lines of inflection are thus {\em mesoscopic}. They contain large numbers of atoms (microscopic units) and are amenable to a continuum description, but are yet much smaller than the typical size of the sheet. Arguments from energy minimization, while implying the existence of these higher order defects, do not address the question of their evolution. One has to necessarily go beyond the elastic energy~\eqref{eq:elastic} and incorporate dissipative effects that  are crucial in determining a thermo-mechanically consistent description of the coupled evolution of the shape $y:\Omega \to \mathbb{R}^3$ and the internal geometry, given by the tensors $g$ and $b$.

\section*{Notation}
We define the notation employed in the paper in one place for convenience. 

When a function on a domain is discontinuous across a  (non-planar) surface $S$, we assume that its values along any sequence of points from either side of the surface approaching any fixed point on the surface take on a unique pair of limiting values, each element in the pair corresponding to the limit from one side. The difference of these limiting values, one for each point on the surface, is defined as the jump (denoted by $\llbracket \cdot \rrbracket$) of the function on the surface. If $\nu(x)$ is the unit normal to $S$ at $x \in S$, we say that $x^\pm$ is a point on the $\pm$ side of $S$ at $x$ depending on $(x^\pm - x) \cdot \nu(x) \gtrless 0$, respectively.

We think of an $n^{th}$ order tensor as a linear transformation between the space of vectors (in the translation space of three-dimensional Euclidean space, also $1^{st}$-order tensors) to the space of $(n-1)^{th}$-order tensors, with its transpose defined in the natural way as being a linear transformation from the space of $(n-1)^{th}$-tensors to the space of vectors. All tensors components will be written w.r.t. the basis, $(e_1, e_2, e_3)$ of a fixed Rectangular Cartesian coordinate system and all partial derivatives, denoted often by a subscript comma, will be w.r.t coordinates of this system. The Einstein summation convention will be used unless otherwise stated. Superposed dots will represent partial derivatives w.r.t. time. If $A$ is a $p^{th}$-order tensor then the operators $\nabla$, $div$, $curl$ may be defined as
\begin{align}
    \nabla A & = A_{i_1 \dots i_p,k}\  e_{i_1} \otimes \ldots \otimes e_{i_p} \otimes e_k \notag\\
    div \, A & = A_{i_1 \dots i_{p-1}k,k}\  e_{i_1} \otimes \ldots \otimes e_{i_{p-1}} \notag\\
    curl \, A & = e_{kr i_p} A_{i_1 \dots i_p,r}\  e_{i_1} \otimes \ldots \otimes e_{i_{p-1}} \otimes e_k, \notag
\end{align}
(with invariant meaning independent of the choice of coordinate system and its basis, of course). The range of all indices above is $1$ to $3$ and $e_{ijk}$ represents a component of the third-order alternating tensor.

The symbol $\inprod{i}$ represents a contraction on $i$ indices between two tensors. For any tensor $A$, we define the tensor obtained by symmetrizing in the first two indices as $A^{(s)}$ and the one obtained by antisymmetrizing in the first two indices from the left as $A^{(a)}$. We denote the deviatoric part of a second-order tensor by the superscript $dev$.
\section{Motivation for kinematics of the theory}\label{sec:motiv}
In this section we provide some intuition on the defect kinematics we adopt for our theory of branch point singularities. This is first done by explicitly constructing a continuously differentiable deformation of a non-simply connected domain whose second derivative has a \emph{prescribed, constant jump} across a planar surface in the body. 
\begin{figure}
\centering
\includegraphics[width=4.5in, height=4in]{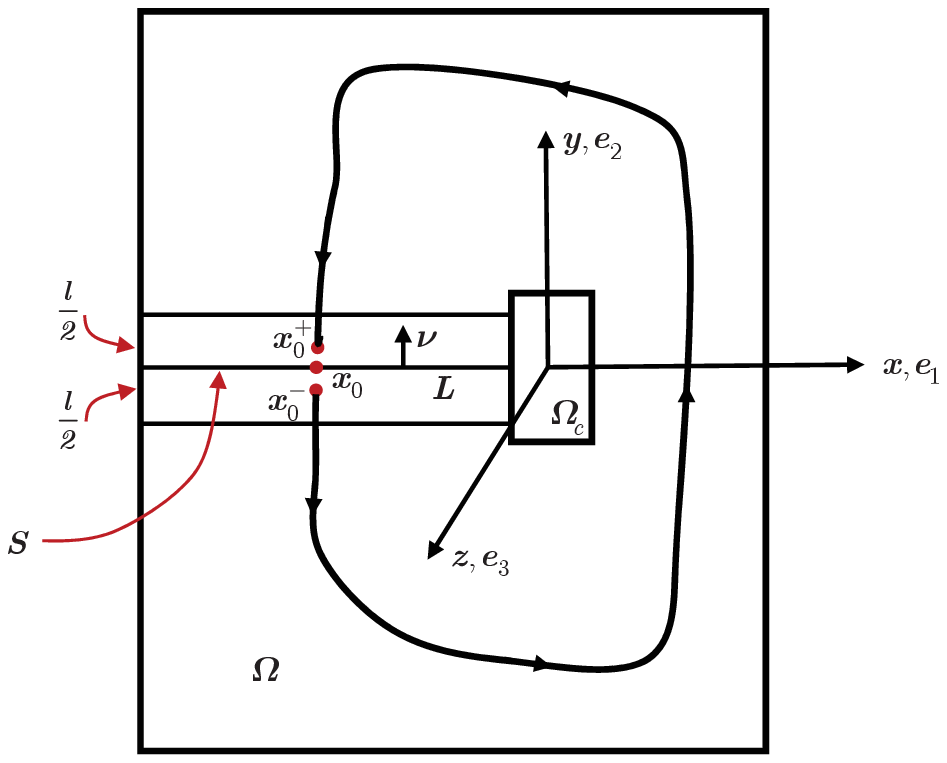}
\caption{Schematic of set up.}
\label{fig:domain}
\end{figure}

With reference to Fig. \ref{fig:domain}, we think of $\Omega$ occupying a simply connected $d= 2 \mbox \ {or} \ 3$-dimensional domain of ambient Euclidean space. Here, it may be viewed either as a right-cylinder ($d=3$) or a cross-section perpendicular to its axis ($d = 2$). We choose a rectangular Cartesian coordinate system with the $z$-axis as the axis of the cylinder; $e_i, i=1,2,3$ are the unit vectors along the $x,y,z$ directions, respectively. $\Omega_c$ is a cylindrical subset of $\Omega$ with rectangular cross-section centered on the $z$-axis. The region $\Omega_h := \Omega \backslash \Omega_c$ is not simply-connected. $S$ is a surface in $\Omega_h$ such that $ D := \Omega_h \backslash S$ is simply connected. The layer $L$ is defined as $L = \left\{ (x,y,z) \in \Omega_h  \, \vert \, x < 0, - \frac{l}{2} \leq y \leq \frac{l}{2} \right\}$ and the surface $S = \left\{ (x,y,z) \in \Omega_h \, \vert \, x < 0, y = 0 \right\}$. We will refer to $\Omega_c$ as a \emph{core}.

Our \emph{goal} in this section is to construct a vector field $\tilde{y}^{(l)} : \Omega_h \rightarrow \mathbb{R}^n$, $n \in \mathbb{N}$,  $n \geq d$, $0 < l \in \mathbb{R}$, with $\tilde{y}^{(l)} \in C^1(\Omega_h)$ and the jump in $\nabla^2 \tilde{y}^{(l)}$ across $S$ a specified constant, with the jump blowing up as $l \rightarrow 0$ maintaining $\lim_{l \rightarrow 0} \nabla \tilde{y}^{(l)} \in C^0(\Omega_h)$.

A necessary condition for $\tilde{y}^{(l)} \in C^1(\Omega_h)$ is that the jump in its second derivative across $S$ be of the form $\left\llbracket \nabla^2 \tilde{y}^{(l)} \right\rrbracket = A \otimes \nu$, where $\nu$ is the unit normal field on $S$ (with arbitrarily chosen orientation) and $A$ is a $\mathbb{R}^{n \times d}$ valued matrix field on $S$. Noting that $l^{-1} \left\llbracket \nabla^2 \tilde{y}^{(l)} \right\rrbracket$ may be formally considered an approximate discrete directional derivative of $\nabla^2 \tilde{y}^{(l)}$ in the direction $\nu$ (if the discontinuity were ignored), we define the field
\begin{equation}\label{eqn:Z}
Z := 
\begin{cases}
\frac{1}{l} A \otimes \nu \otimes \nu \qquad & \mbox{in} \ L\\
0 & \mbox{in} \ \Omega_h \backslash L
\end{cases}
\end{equation}
with $A$ and $\nu = e_2$ constants, and seek to construct solutions to the equations
\begin{equation}\label{eqn:gov_WY}
\begin{rcases}
\nabla W & = Y\\
\nabla Y & = \left. Z \right\vert_D
\end{rcases} \qquad \mbox{in} \  D.
\end{equation}
The restriction of $Z$ to $D$ is used since, even though $Z$ is (distributionally) $curl$-free in $\Omega_h$ (we interpret the $curl$ of a matrix field as row-wise $curl$s), $\Omega_h$ is not simply connected but $D$ is and hence we are guaranteed a solution $Y$ in $D$, unique up to a constant.

For any such $Y$ field, we note that $curl \,Y = 0$ in $D$ by the symmetry in the last two entries of $Z$, i.e. $(\nabla Y e_l)e_k - (\nabla Y e_k)e_l = (Z e_l)e_k - (Z e_k)e_l = 0$. Thus $W$ satisfying \eqref{eqn:gov_WY} can be constructed, and $W$ is also unique in $D$ up to a constant for a given $Y$.

Arbitrarily fix one of the available $Y$ fields. Such a $Y$ has the explicit representation
\begin{equation*}
Y(x; x_0) = \lim_{x_0^- \to x_0} \left( Y \left(x_0^-\right) + \int_{x_0^-}^x Z \, dx \right), \qquad x \in D,
\end{equation*}
fopr $x_0$ being \emph{any} point on the surface $S$, and the line integral is along any path from $x_0^-$ to $x$ contained in $D$. Now choose any path going from $x_0^-$ to $x_0^+$ (see Fig. \ref{fig:domain}) contained in $D$ with the stipulation that it go through the points $x_0 \pm \frac{l}{2} e_2$ and the segments between $x_0$ and $x_0 \pm \frac{l}{2} e_2$, respectively, are parallel to $e_2$. Then, along the segment $x(s) = x_0^- - (s-s^-) e_2, 0 < s^- \leq s \leq \frac{l}{2}$,
\begin{equation}\label{eqn:Y_bot_layer}
Y\left(x_0 - \frac{l}{2} e_2 \right) = Y(x_0^-) + \bigintss_{s^-}^{\frac{l}{2}} \frac{A \otimes \nu}{l} ( e_2 \cdot -e_2)\, ds = Y(x_0^-) - \frac{A \otimes \nu}{l} \left( \frac{l}{2} - s^- \right).
\end{equation}
$Y(x(s))$ remains constant along the path between $x_0 - \frac{l}{2} e_2$ and $x_0 + \frac{l}{2} e_2$. Therefore, using the segment $x(s) = x_0 + \frac{l}{2} e_2 - s e_2, 0 \leq s \leq \left( \frac{l}{2} - s^+ \right)$ we have
\begin{equation*}
\begin{split}
Y(x_0^+) & =  Y\left( x_0 + \frac{l}{2} e_2 \right) + \int_{x_0 + \frac{l}{2} e_2}^{x_0^+} Z \, dx\\
& = Y\left( x_0 - \frac{l}{2} e_2 \right) - \frac{A \otimes \nu}{l} \left( \frac{l}{2} - s^+ \right)\\
& = Y(x_0^-) - \frac{A \otimes \nu}{l} \left( \frac{l}{2} - s^- \right) - \frac{A \otimes \nu}{l} \left( \frac{l}{2} - s^+ \right).
\end{split}
\end{equation*}
Hence, the jump in $Y$ at $x_0$ is given by
\begin{equation}\label{eqn:jumpY1}
\llbracket Y \rrbracket (x_0) = \lim_{\substack{x_0^\pm \to x_0 \\ s^\pm \to \, 0}} Y (x_0^+) - Y(x_0^-) = - A \otimes \nu.
\end{equation}
Since $x_0 \in S$ and $Y$ such that $\nabla Y = Z$ in $D$ were chosen arbitrarily, \eqref{eqn:jumpY1} holds for all $x_0 \in S$ and admissible $Y$ in the specified class. Thus $\llbracket Y \rrbracket$ is unique in that class, independent of position on $S$, and given by the constant $- A \otimes \nu$.

We note that $Y^*:\Omega_h \rightarrow \mathbb{R}^{n \times d \times d}$ may be viewed as a discontinuous function with the specification
\begin{equation*}
Y^*(x) =
\begin{cases}
\lim\limits_{\,x^- \to\, x} Y(x^-) - \half A \otimes \nu, &\qquad  x \in S\\
Y(x), & \qquad x \in D.
\end{cases}
\end{equation*}
where the points $x^- \in D $ belong to the $-$ side of $S$ at $x$.

We now evaluate the jump in the field $W$ on $S$. 

As already observed, for any $Y$ satisfying \eqref{eqn:gov_WY} a $W$ field in $D$ can also be constructed and this will have the representation
\[
W(x;y) = W(y) + \int_y^x Y \, dx, \qquad x,y \in D,
\]
for any path linking $y$ to $x$ in $D$. We now arbitrarily fix an admissible field $Y$ and choose the same path from $x_0^-$ to $x_0^+$ used in deducing its jump on $S$.

Along $x(s) = x_0^- - (s - s^-) e_2$, $s^- \leq s \leq \frac{l}{2}$, $Y(s) = Y(x_0^-) - \frac{A \otimes \nu}{l} (s - s^-)$ and 
\begin{equation}\label{eqn:W_bot_layer}
\begin{split}
W \left( x_0 - \frac{l}{2} e_2 \right) & = W(x_0^-) + \bigintss_{s^-}^{\frac{l}{2}} \left[ \frac{-A \otimes \nu}{l} (s - s^-) \right] (- e_2)\, ds + Y(x_0^-) \int_{s^-}^{\frac{l}{2}} (-e_2) \, ds\\
& = W (x_0^-) + \bigintss_0^{\frac{l}{2} - s^-} \frac{A \otimes e_2}{l} (e_2) s' \, ds' - \left( \frac{l}{2} - s^- \right) Y(x_0^-) e_2\\
& = W (x_0^-) + \frac{A}{2l} \left(\frac{l}{2} - s^- \right)^2 - \left(\frac{l}{2} - s^- \right) Y(x_0^-) e_2.
\end{split}
\end{equation}
Since $Y$ remains constant at the value given by \eqref{eqn:Y_bot_layer} along the chosen path from $x_0 - \frac{l}{2} e_2$ to $x_0 + \frac{l}{2} e_2$,
\begin{equation}\label{eqn:W_top_layer}
\begin{split}
W \left( x_0 + \frac{l}{2} e_2 \right) & = W \left( x_0 - \frac{l}{2} e_2 \right) + \bigintsss_{x_0 - \frac{l}{2} e_2}^{x_0 + \frac{l}{2} e_2} Y \, dx\\
& = W \left( x_0 - \frac{l}{2} e_2 \right) + l\, Y\left( x_0 - \frac{l}{2} e_2 \right)\, e_2\\
& = W(x_0^-)  + \frac{A}{2l} \left(\frac{l}{2} - s^- \right)^2 - \left(\frac{l}{2} - s^- \right) Y(x_0^-) e_2 + l\, Y(x_0^-) e_2 - \left( \frac{l}{2} - s^- \right) A.
\end{split}
\end{equation}
using \eqref{eqn:Y_bot_layer} and \eqref{eqn:W_bot_layer}. Now
\begin{equation}\label{eqn:W_S+}
W (x_0^+) = W \left( x_0 + \frac{l}{2} e_2 \right) + \bigintsss_{x_0 + \frac{l}{2} e_2}^{x_0^+} Y \, dx
\end{equation}
and $Y(s)$ along the segment $x(s) = x_0 + \frac{l}{2} e_2 - s e_2, 0 \leq s \leq \frac{l}{2} - s^+ $ is given by
\[
Y(s) = Y \left( x_0 + \frac{l}{2} e_2 \right) + \int_0^s Z (s) (- e_2) \, ds = Y\left( x_0 + \frac{l}{2} e_2 \right) - s\frac{A \otimes \nu}{l},
\]
so that
\begin{equation*}
\begin{split}
\bigintsss_{x_0 + \frac{l}{2} e_2}^{x_0^+} Y \, dx & = \bigintss_0^{\frac{l}{2} - s^+} \left[ Y \left( x_0 + \frac{l}{2} e_2 \right)  - s\frac{A \otimes \nu}{l} \right] (- e_2) \,ds\\
& = \left[- Y \left( x_0 + \frac{l}{2} e_2 \right) e_2 \right] \left( \frac{l}{2} - s^+ \right) + \frac{A}{2l} \left( \frac{l}{2} - s^+ \right)^2,
\end{split}
\end{equation*}
and therefore \eqref{eqn:W_top_layer}, \eqref{eqn:W_S+}, and \eqref{eqn:Y_bot_layer}, noting $Y \left( x_0 - \frac{l}{2} e_2 \right) = Y \left( x_0 + \frac{l}{2} e_2 \right)$, imply
\begin{equation*}
\begin{split}
W (x_0^+) - W(x_0^-) =\  & \frac{A}{2l} \left(\frac{l}{2} - s^- \right)^2 - \left(\frac{l}{2} - s^- \right) Y(x_0^-) e_2 + l\, Y(x_0^-) e_2 - \left( \frac{l}{2} - s^- \right) A \\
& + \left[- \left\{ Y(x_0^-) - \frac{A \otimes \nu}{l} \left( \frac{l}{2} - s^- \right) \right\} e_2 \right] \left( \frac{l}{2} - s^+ \right) + \frac{A}{2l} \left( \frac{l}{2} - s^+ \right)^2.
\end{split}
\end{equation*}
Thus,
\begin{equation}\label{eqn:Wjump}
\llbracket W \rrbracket (x_0) = \lim_{\substack{x_0^\pm \to\, x_0\\ s^\pm \to \,0}} W (x_0^+) - W(x_0^-)  = \frac{A}{2l} \left(\frac{l}{2} \right)^2 - \frac{l}{2} A + \frac{A}{l} \left(\frac{l}{2} \right)^2  + \frac{A}{2l} \left(\frac{l}{2} \right)^2 = 0.
\end{equation}

We now define the function $W^*: \Omega_h \rightarrow \mathbb{R}^{n \times d}$ as
\begin{equation}\label{eqn:W*}
W^*(x) =
\begin{cases}
\lim\limits_{x^- \to \, x} W(x^-) = \lim\limits_{x^+ \to \, x} W(x^+), &\qquad x^\pm \in D, x \in S\\
W(x), & \qquad x \in D,
\end{cases}
\end{equation}
where the points $x^\pm$ belong to the $\pm$ sides of $S$ at $x$, respectively. $W^*$ is a continuous function on $\Omega_h$.

We now assume that the constant $A$ is of the form $A = a \otimes \nu$ for $a \in \mathbb{R}^n$. Then, in $D$, $\nabla^2 W = Z = a \otimes \nu \otimes \nu \otimes \nu$ so that $\nabla W = (\nu \cdot x) a \otimes \nu \otimes \nu  + C$ where $C \in \mathbb{R}^{n \times d \times d}$ is a constant. This constant is free to choose, without loss of generality (related to the choice of $Y(x_0^-)$, for instance), and we assume that it satisfies $(C e_i)e_j = (C e_j) e_i$ for $i, j = 1, \ldots,d$. Then $curl \, W = curl \, W^* = 0$ in $D$. This further implies that the line integral $\int W^* \, dx = b \in \mathbb{R}^n$ is a constant for any closed contour encircling $\Omega_c$.

If $b = 0$, then we define 
\[
\widetilde{W} = W^* \ \mbox{in} \ \Omega_h.
\]
If not, we explicitly solve the system
\begin{equation}\label{eqn:RG}
curl \, \widehat{W} = - b \otimes e_3\, \delta_{z-axis} =: \widehat{\alpha} \qquad \mbox{in} \ \Omega.
\end{equation}
Solutions exist to this system (e.g. an explicit solution on star-shaped domains can be written down by using the Riemann-Graves integral operator \cite{edelen2005applied}) that belong to  $C^1(\Omega_h)$. Forcing by the Dirac distribution is not necessary; functions of $(x, y)$ with support in a cylinder contained in $\Omega_c$ satisfying $\int_A \hat{\alpha}\, e_3 \, da = -b$ for any area patch $A$ threaded by the cylinder also suffice for generating such solutions \cite{acharya2001model}). Then defining
\[
\widetilde{W} = \left. \widehat{W}\right|_{\Omega_h} + W^*  \ \mbox{in} \ \Omega_h,
\]
we note that $\bigintsss \widetilde{W} \, dx = 0$ for any closed contour encircling $\Omega_c$ and that $\widetilde{W} \in C^0(\Omega_h)$. Then we define $\tilde{y}: \Omega_h \rightarrow \mathbb{R}^n$ by
\begin{equation}\label{eqn:ytide}
\tilde{y} (x; z) = p + \int_z^x \widetilde{W} \, dx, \qquad x,z \in \Omega_h
\end{equation}
for arbitrarily fixed $z \in \Omega_h$ and a constant $p \in \mathbb{R}^n$. 

Clearly, $\tilde{y}$ satisfies $\nabla \tilde{y} = \widetilde{W}$ on $\Omega_h$ and $\tilde{y} \in C^1(\Omega_h)$.

Consider the constant vector $a \in \mathbb{R}^n$ to be parametrized by the layer width $l$ as
\[
a^{(l)} = \gamma \, l^{\beta - 1}, \qquad \gamma \in \mathbb{R}^n, 0 < \beta \in \mathbb{R}.
\]
All fields constructed with the use of $A = a^{(l)} \otimes \nu$ are denoted by a superscript $(l)$. We have the following properties:
\begin{itemize}
\item For $l \to 0$, $0 < \beta < 1$, $\tilde{y}^{(l)} \in C^1(\Omega_h)$, $\left \vert\nabla^2 \tilde{y}^{(l)} \right \vert \to \infty$ in $\Omega_h$ since $\lim_{x^+ \to x} Y(x^+) = \lim_{x^- \to x} Y(x^-) + \left( l^{\beta -1} \right) \gamma \otimes \nu \otimes \nu$ on $S$. For $\beta = 0$, $W^{(l)}$ fails to remain continuous as $l \to 0$ \eqref{eqn:W_bot_layer}, providing an obstruction for $\tilde{y}^{(l)}$ to belong to $C^1(\Omega_h)$.
\item For $l \to 0$, $\beta = 1$, $\tilde{y}^{(l)} \in C^1(\Omega_h)$, $\nabla^2 \tilde{y}^{(l)} \in C^0(D)$ and $\left\llbracket \nabla^2 \tilde{y}^{(l)} \right \rrbracket$ is bounded on $S$. This conclusion also holds for any value of $\beta \geq 0$ when $l >0$ is held fixed.
\item For $l \to 0$, $\beta > 1$,  $\tilde{y}^{(l)} \in C^2(\Omega_h)$.
\end{itemize}
\begin{remark}
While the above considerations have dealt with one singular surface, the linearity of the construct on the prescribed field $Z$ makes it clear that exactly similar arguments hold for the superposition of a set of deformations, each element of which contains a single planar surface of discontinuity of arbitrary orientation in $\Omega_h$ terminating on $\Omega_c$. Considering $y^i$, $i = 1 \, \emph{to} \, n \in \mathbb{Z}^+$, each corresponding to a specified $Z^i$ field,
the composite, superposed deformation $ \sum_{i=1}^n y^i$ is $C^1(\Omega_h)$, with generally discontinuous second derivatives on each of the $S^i$ corresponding to the specified $Z^{i}$ field. This corresponds to situations with a single branch point \cite{GV2011,GV2012,GV2013} as exemplified by the piecewise quadratic monkey-saddle that we discussed in Sec.~\ref{sec:branch-points}.

Furthermore, given a fixed, simply connected domain $\Omega$, let $\Omega_c^i \subset \Omega$, $ i = 1 \, \emph{to} \, n$, be a set of non-intersecting cores with $\Omega_h^i := \Omega \backslash \Omega_c^i$. Let each $Z^i$ now be specified on the domain $\Omega^i_h$. Then each $y^i$ is $C^1(\Omega_h^i)$. Thus, $ \sum_{i=1}^n y^i \in C^1 \!\left(\cap_{i=1}^n \Omega_h^i \right)$. This corresponds to configurations with multiple branch-points \cite{EPL_2016,toby_todo}.


\end{remark}
\begin{remark}
For thin objects modeled by $d = 2$, the construction above is a representation of folds without ridges. In Sections \ref{sec:kin} and \ref{sec:thermo} we develop a continuum mechanical theory that encompasses the mechanics of such folds in simply connected domains within a setting that allows for deformations with less smoothness.
\end{remark}
\begin{remark}\label{rem:defect_stress}
Consider $d = 2, n = 2$ and $b \neq 0$, and assume that $W^*(x), x \in \Omega_h$, is invertible. A field $y^*:D \to \mathbb{R}^2$ satisfying $\nabla y^* = W^*$ in $D$ can be constructed that may be interpreted as a discontinuous deformation of $\Omega_h$. Now consider the metric $g := W^{*T}W^{*}$ on $\Omega_h$. By the Nash $C^1$ embedding theorem, there exists a $C^1$ deformation 
$z:\Omega_h \to \mathbb{R}^3$ with $ (\nabla z)^T \nabla z = g = (\nabla y^*)^T \nabla y^*$.

For a mechanistic interpretation, consider the configuration in $\mathbb{R}^3$ defined by $z(\Omega_h)$ as the stress-free, global reference configuration in a higher dimensional space $(\mathbb{R}^3)$ corresponding to a stressed body with a dislocation (with excluded core) in $\mathbb{R}^2$ represented by $\Omega_h$. The stress-free reference cannot be represented by a compatible mapping of $\Omega_h$ in  the lower-dimensional space $\mathbb{R}^2$; instead, one of its stress-free representations in $\mathbb{R}^2$ is defined by the configuration $y^*(\Omega_h)$. The stress-producing elastic Right-Cauchy Green tensor field is given by $(W^{*-1})^T W^{*-1}$ on $\Omega_h$.
\end{remark}
\subsection{The discontinuity of the deformation of a non simply connected domain with prescribed third `deformation gradient'}\label{sec:wein}
Consider the domain $\Omega_h$ of Fig. \ref{fig:domain} which is rendered simply connected by a single cut-surface $S$ which is not necessarily planar. As before, we refer to $\Omega_h \backslash S =:D$. We consider $Z: \Omega_h \to \mathbb{R}^{n \times d \times d \times d}$ as a given field for which $((Z (x) e_l)e_k)e_j$ is invariant w.r.t interchanges of $e_j, e_k, e_l$ for any values of $j,k,l \in \{1, \dots, d\}$. Furthermore, we assume that $Z \in C^0(\Omega_h)$, and $curl \, Z = 0$ in $\Omega_h$. We are now interested in the construction of a field $y: D \to \mathbb{R}^n$ that satisfies
\[
\nabla^3 y = Z
\]
and characterizing the jump field $\llbracket y \rrbracket$ on $S$.

Define, for $x \in \Omega_h$, $[((Z (x) e_l)e_k)e_j]\cdot E_I =: Z_{Ijkl}(x)$, $I = 1,\dots, n$ and $j,k,l = 1,\ldots,d$, where $E_I$ represents an element of an orthonormal basis in $\mathbb{R}^n$. $Z_{Ijkl}$ is symmetric in the indices $j,k,l$. Now construct $Y:D \to \mathbb{R}^{n \times d \times d}$ satisfying
\begin{equation}\label{eqn:Y}
\frac{\partial Y_{Ijk}}{\partial x_l} = Z_{Ijkl},
\end{equation}
which is possible since $curl Z = 0$ and $D$ being simply connected. We note that $Y_{Ijk} (x) - Y_{Ikj}(x) = Y_{Ijk} (y) - Y_{Ikj}(y)$ for $x,y \in D$, due to the symmetry of $Z_{Ijkl}$ in $j,k$ and the connectedness of $D$. Since the construction of $Y$ allows the free specification of its value at one point of $D$, it can be assumed without loss of generality that $Y_{Ijk} = Y_{Ikj}$ in $D$.

Equation \eqref{eqn:Y} and the symmetry of $Z$ in the last two indices imply $curl\, Y = 0$ in $D$. Thus it is also possible to construct $W : D \to \mathbb{R}^{n \times d}$ satisfying
\begin{equation}\label{eqn:W}
\frac{\partial W_{Ij}}{\partial x_k} = Y_{Ijk}.
\end{equation}
Furthermore, \eqref{eqn:W} and the symmetry of $Y$ in its last two indices imply that a function $y: D \to \mathbb{R}^n$ can be constructed satisfying
\begin{equation}\label{eqn:y}
\frac{\partial y_{I}}{\partial x_j} = W_{Ij}.
\end{equation}

Now, because $Z$ is $curl$-free in $\Omega_h$, we have by Stokes' theorem that
\begin{equation}\label{eqn:Gamma}
\int Z\, dx =: \Gamma \in \mathbb{R}^{n \times d \times d} \ \mbox{a constant, for the line integral over \emph{any} closed loop encircling} \  \Omega_c.
\end{equation}
By \eqref{eqn:Y} and \eqref{eqn:W}, this further implies that
\begin{equation}\label{eqn:jumpY}
\Gamma = \llbracket Y \rrbracket (x) = \llbracket \nabla W \rrbracket (x), \qquad x \in S.
\end{equation}
Let $x_0, x \in S$ be connected by a curve $c$ contained in $S$. Consider curves $c^+$ and $c^-$ on the $\pm$ sides of $S$ connecting $x_0^\pm$ to $x^\pm$. Then
\begin{equation}\label{eqn:jumpW}
W(x^\pm) = W(x_0^\pm) + \int_{x_0^\pm}^{x^\pm} \nabla W (c^\pm)\, dc^\pm \implies \llbracket W \rrbracket (x) = \llbracket W \rrbracket (x_0) + \Gamma ( x - x_0)
\end{equation}
as $c^\pm \to c$. Similarly, \eqref{eqn:y} implies
\begin{equation}\label{eqn:jumpy_pre}
\begin{split}
y(x^\pm) & = y(x_0^\pm) + \int_{x_0^\pm}^{x^\pm} \nabla y (c^\pm)\, dc^\pm\\
\implies \llbracket y \rrbracket (x) & = \llbracket y\rrbracket (x_0) + \int_{x_0}^x \llbracket W \rrbracket (c) \, dc = \llbracket y\rrbracket (x_0) + \int_{x_0}^x \Big\{ \llbracket W \rrbracket (x_0) + \Gamma (c - x_0) \Big\} \, dc\\
& = \llbracket y\rrbracket (x_0) + \Big( \llbracket W \rrbracket (x_0)\Big) (x - x_0) + \int_0^{x - x_0} \Gamma c' \, dc'
\end{split}
\end{equation}
Now, due to the symmetry of $\Gamma$ in its last two indices, $\Gamma_{Ijk} c'_k \frac{dc'_j}{ds} = \frac{1}{2} \frac{d}{ds} (\Gamma_{Ijk} c'_k c'_j)$ and the last line integral in \eqref{eqn:jumpy_pre} evaluates to $\frac{1}{2} \left( \Gamma (x - x_0) \right) (x - x_0)$ so that \eqref{eqn:jumpy_pre} implies
\begin{equation}\label{eqn:jumpy}
\llbracket y \rrbracket (x) = \llbracket y\rrbracket (x_0) + \Big( \llbracket W \rrbracket (x_0)\Big) \inprod{1} (x - x_0) + \half \, \Gamma \inprod{2} \left[ (x - x_0) \otimes (x - x_0) \right], \qquad \forall x,x_0 \in S.
\end{equation}

\begin{remark}
The jump in the deformation $y$ across the cut-surface $S$ is not arbitrary, being characterized by a finite set of parameters. One choice for this parameter set is the jump of the deformation at an arbitrarily fixed point on $S$, the jump of $W$ at the same point, and $\Gamma$, the latter being a constant decided by the given field $Z$.
\end{remark}
\begin{remark}
$\llbracket W \rrbracket$ is not constant on $S$ even though $curl \, Y = 0$ in $D$ 
unless the vector joining any two points on $S$ lies in the null-space of $\Gamma$ by \eqref{eqn:jumpW}. For $S$ a planar surface with unit normal $\nu$ and $\Gamma$ of the form $a \otimes \nu \otimes \nu$, $a \in \mathbb{R}^n, \nu \in \mathbb{R}^d$ constants, \eqref{eqn:jumpW} implies that $\llbracket W \rrbracket$ is constant on $S$. If, moreover $\llbracket W \rrbracket - (\llbracket W \rrbracket \nu) \otimes \nu = 0$, then $\llbracket y \rrbracket$ is also a constant on $S$. These are all conditions satisfied by the example worked out in the preamble of this Section.
\end{remark}
\begin{remark}
The argument remains unchanged for the case $\Omega_h$ is just a punctured domain, i.e. $\Omega_c$ shrinks to a point (a curve).
\end{remark}
\begin{remark}
The result \eqref{eqn:jumpy} is an extension of Weingarten's theorem \cite{wein, delph_wein, volt, delph_volt} and the Weingarten theorem for g.disclinations \cite{zhang2018relevance}.
\end{remark}

\section{Kinematics}\label{sec:kin}
In this section we propose the kinematics for a model of the type of discontinuities treated in Sec. \ref{sec:motiv}, to be broadly applied to the mechanics of materials. For that purpose, it is essential to deal with simply connected, compact domains containing the said discontinuities. The excluded core regions are now included in the domain as are the excluded surfaces of discontinuity. {\em Roughly speaking}, we consider an additive split of fields into `regular' and `singular' parts whenever the field in question contains high magnitudes concentrated in `thin' regions approximating smooth lower-dimensional $(< d)$ sets; the support of the singular part of the field contains these regions of high concentration and that of the regular part contains the support of the rest of the field, including regions supporting approximate discontinuities. Importantly, both the singular and regular parts are assumed to be at least integrable functions as we want to write governing equations for these fields in the form of pde that can at least be made sense of in some weak manner. Thus, we take a somewhat microscopic point of view, assuming that discontinuities and singularities of certain fields when viewed from a macroscopic scale have a smoother definition at a microscopic scale that we describe by additional `eigenwall' fields. We also adopt the point of view that once macroscopic theories generate discontinuities and singularities, in most circumstances additional physical insight beyond the constraints placed by the governing equations of the macroscopic theory are required to define evolution with a modicum of uniqueness. We develop such a model in the rest of the paper.

We refer to a fixed \emph{reference} configuration, a simply connected compact region as $B$. In terms of the \emph{displacement} field $u$ and the $i$-\emph{eigenwall} fields $S^{(i)}, i \in \{1,2,3\}$, we define the \emph{i-elastic distortions} $Y^{(i)}, i \in \{0,\ldots,4\}$, as
\begin{equation}\label{defY}
\begin{split}
Y^{(4)} &:= \nabla Y^{(3)}\\
Y^{(i)} & := \nabla Y^{(i-1)} - S^{(i)} \qquad i \in \{1,2,3\}\\
Y^{(0)} & := u.
\end{split}
\end{equation}
($Y^{(0)}$ is analogous to the field $y$ of Sec. \ref{sec:motiv}, $Y^{(1)}$ to $W$, $Y^{(2)}$ to $Y$, and $Y^{(3)}$ to $Z$). Thus $Y^{(0)} = u$ and $Y^{(4)}$, the gradient of the regular part of the gradient of the 3-elastic distortion, are assumed to have no `singular' parts. We now define the \emph{`composite' eigenwall} fields $\widehat{S}^{(i)}, i = 1,2,3$, as
\begin{equation}\label{defShat}
\begin{split}
\Ythree & = \nabla \Ytwo - \Sthree = \nabla^3 u - \Shatthree; \qquad \Shatthree := \nabla^2 \Sone + \nabla \Stwo + \Sthree\\
\Ytwo & = \nabla \Yone - \Stwo = \nabla^2 u - \Shattwo; \qquad \Shattwo := \nabla \Sone + \Stwo\\
\Yone & = \nabla Y^{(0)} - \Sone = \nabla u - \Shatone;\  \qquad  \Shatone := \Sone,
\end{split}
\end{equation}
and we note that
\begin{equation}\label{eqn:S_Shat}
S^{(i)} = \widehat{S}^{(i)} - \nabla \widehat{S}^{(i-1)} \qquad i \in \{1,2,3\}.
\end{equation}

Physical considerations related to predicting stress fields of terminating twin boundaries and the stress-free, compatible, elastic, twinning shear distortions of through-twin boundaries \cite{zhang2018finite} motivate the introduction of the following Stokes-Helmholtz (SH) decompositions:
\begin{equation}\label{eqn:SH_S}
\begin{split}
\begin{rcases}
 S^{(i)} = \nabla H^{(i)} - \chi^{(i)}& \\
 curl \, \chi^{(i)} = -curl \, S^{(i)} &\\
 div \, \chi^{(i)} = 0 &\\
 div \, \nabla H^{(i)} = div \, S^{(i)} &\\
\end{rcases}&
\qquad x\in B, \qquad i \in \{1,2,3\},\\
\begin{rcases}
\chi^{(i)} n = 0 & \\
 \nabla H^{(i)} n = S^{(i)} n &\\ 
 \end{rcases}&
\qquad x\in \partial B, \qquad i \in \{1,2,3\}.
\end{split}
\end{equation}
We will also consider exactly analogous SH decompositions for the fields
\begin{equation}\label{eqn:SH_Shat}
\widehat{S}^{(i)} = \nabla \widehat{H}^{(i)} - \widehat{\chi}^{(i)}, \qquad i \in \{1,2,3\}.
\end{equation}
Combining \eqref{eqn:S_Shat} and \eqref{eqn:SH_Shat} and noting the uniqueness of the SH decomposition we have
\begin{equation}\label{eqn:Hhat_H}
H^{(i)} = \widehat{H}^{(i)} - \widehat{S}^{(i-1)}, \qquad i \in \{1,2,3\},
\end{equation}
up to at most a spatially constant function of time which we will assume to be a time-independent constant.
Defining
\begin{equation}\label{eqn:Yhat_Y}
\widehat{Y}^{(i)} := Y^{(i)} - H^{(i+1)}, \qquad i \in \{1,2,3\}
\end{equation}
(noting that $H^{(4)} = 0$), we define the $i$-\emph{defect density} tensors for $i \in \{1,2,3\}$ from \eqref{defY} and \eqref{eqn:Yhat_Y} as
\begin{equation}\label{eqn:defalpha}
\begin{split}
& \alpha^{(i)} := - Y^{(i+1)} \inprod{2}  X = curl \, Y^{(i)} + S^{(i+1)} \inprod{2} X = curl \, \widehat{Y}^{(i)} - \chi^{(i+1)} \inprod{2} X\\
& \widehat{\alpha}^{(i)} := \alpha^{(i)} - S^{(i+1)} \inprod{2} X = curl \, Y^{(i)} = - curl \, S^{(i)} = - curl \, \widehat{S}^{(i)}
\end{split}
\end{equation}
using \eqref{eqn:S_Shat} and $S^{(4)} = \chi^{(4)} = 0$.

Since $\widehat{\alpha}^{(i)}$ are defined locally as a $curl$, the local forms of the conservation laws for topological charge content, $ \int_\Sigma \widehat{\alpha}^{(i)} n\, da$, of an arbitrary area patch $\Sigma$ is given by 
\begin{equation}\label{eqn:conserv_law}
\dot{\overline{{\widehat{\alpha}}^{(i)}}} = -curl \left( \widehat{\alpha}^{(i)} \times V^{\parallel(i)} \right), \qquad i \in \{1,2,3\}
\end{equation}
where $V^{\parallel(i)}$, for each $i$, is a vector field. $V^{\parallel(i)}$ is the velocity field of the $i$-defect density field. Combining \eqref{eqn:defalpha} and \eqref{eqn:conserv_law},  we have that
\begin{equation}\label{S_evol}
curl\, \left( \dot{\overline{ S^{(i)} } } - \widehat{\alpha}^{(i)} \times V^{\parallel(i)} \right) = 0 \Longleftrightarrow \dot{\overline{ S^{(i)} } } = \left(-curl\, S^{(i)}\right) \times V^{\parallel(i)} + \nabla F^{(i)},  \qquad i \in \{1,2,3\}
\end{equation}
for some $F^{(i)}$ that can be prescribed. Equations \eqref{defShat} and \eqref{S_evol} imply
\begin{equation}\label{Shat_evol}
\dot{ \overline{ \widehat{S}^{(i)}}} = \left(-curl\, S^{(i)}\right) \times V^{\parallel(i)} + \nabla F^{(i)} + \sum_{k=1}^{i-1} \nabla^{i-k} \,\dot{ \overline{S^{(k)}}}, \qquad i \in \{1,2,3\}
\end{equation}
with the last sum vanishing for $i = 1$.

By kinematical arguments related to allowing for transverse motion of walls characterized by localized $S^{(i)}$ fields on surfaces, a part of $F^{(i)}$ is of the form $F^{(i)} = S^{(i)} V^{\perp(i)}$, where $V^{\perp(i)}$ is the velocity of the $i$-eigenwall field. Guided by simplicity in thermodynamic arguments that precludes the appearance of (unremovable) gradients of dislocation and eigenwall velocity fields in the expression for dissipation of the body (see Sec. \ref{sec:thermo}), we make the following choice
\begin{equation}\label{F_i}
\nabla F^{(i)} := \nabla \left(S^{(i)} V^{\perp(i)} \right) - \sum_{k=1}^{i-1} \nabla^{i-k} \,\dot{ \overline{S^{(k)}}}, \qquad i \in \{1,2,3\}.
\end{equation}
In \eqref{S_evol} and \eqref{Shat_evol}, incorporating \eqref{F_i}, $V^{\parallel(i)}$ and $V^{\perp(i)}$ are to be constitutively specified, minimally consistent with the second law of thermodynamics to be globally satisfied for all processes of any body modeled by this theory.

Surfaces of displacement discontinuity (e.g. stacking faults) are not known to move transverse to themselves; moreover, such discontinuitites are often not identifiable based on knowledge of only the current state (and not of the distinguished coherent reference from which displacements are measured). Hence, we will assume $V^{\perp(1)} \equiv 0$. Elastic phase boundaries, i.e. localizations of the $\Sone$ field along surfaces are known to move transverse to themselves, and not much is known about transverse motions of surfaces of discontinuity of the second gradient of elastic distortion, i.e. surfaces of inflection. Thus, we allow $V^{\perp(i)}, i = 2,3$ to be nonvanishing fields in general. Hence, we have the following evolution equations for the eigenwall fields:
\begin{equation}\label{eqn:S_evol_spec}
\begin{split}
\dot{ \overline{{S}^{(1)}}} & = \left(-curl\, S^{(1)}\right) \times V^{\parallel(1)}  = \left(-curl\, \widehat{S}^{(1)}\right) \times V^{\parallel(1)} = \dot{ \overline{ \widehat{S}^{(1)}}}\\
\dot{ \overline{{S}^{(2)}}} + \nabla \, \dot{ \overline{S^{(1)}}} & = \left(-curl\, S^{(2)}\right) \times V^{\parallel(2)} + \nabla \left(S^{(2)} V^{\perp(2)} \right) \\
&  = \left(-curl\, \widehat{S}^{(2)}\right) \times V^{\parallel(2)} + \nabla \left( \left(\Shattwo - \nabla \Shatone \right) V^{\perp(2)} \right) = \dot{ \overline{ \widehat{S}^{(2)}}}\\
\dot{ \overline{{S}^{(3)}}} + \nabla^2 \,\dot{ \overline{S^{(1)}}} + \nabla\, \dot{ \overline{S^{(2)}}} & = \left(-curl\, S^{(3)}\right) \times V^{\parallel(3)} + \nabla \left(S^{(3)} V^{\perp(3)} \right) \\
& = \left(-curl\, \Shatthree \right) \times V^{\parallel(3)} + \nabla \left( \left( \Shatthree - \nabla \Shattwo \right)  V^{\perp(3)} \right)= \dot{ \overline{ \widehat{S}^{(3)}}}\\
\end{split}
\end{equation}
\section{Thermodynamics}\label{sec:thermo}
We assume a free-energy density function of the body with the following dependencies:
\begin{equation}\label{eqn:psi}
\begin{split}
\psi &= \psi^{*} \left(\widehat{Y}^{(1)},\widehat{Y}^{(2)},\widehat{Y}^{(3)},\Shatone,\Shattwo,\Shatthree,\halphaone,\halphatwo,\halphathree, \chi^{(2)}, \chi^{(3)} \right)\\
& = \psi^{**} \left(\Yone,\Ytwo,\Ythree, H^{(2)}, H^{(3)}, \Shatone,\Shattwo,\Shatthree,\halphaone,\halphatwo,\halphathree, \chi^{(2)}, \chi^{(3)}  \right)\\
&= \psi\left(\nabla u, \nabla^2 u, \nabla^3 u, \widehat{H}^{(2)}, \widehat{H}^{(3)}, \Shatone,\Shattwo,\Shatthree,\halphaone,\halphatwo,\halphathree, \chi^{(2)}, \chi^{(3)}  \right),
\end{split}
\end{equation}
using \eqref{eqn:Yhat_Y}, \eqref{defShat}, \eqref{eqn:Hhat_H}, and noting that $H^{(4)} = 0$ (where the argument fields of each of the functions are evaluated at $(x,t)$ to give the value of $\psi(x,t)$). Roughly speaking, the dependencies of $\psi^{*}$ on $\widehat{Y}^{(i)}, \widehat{\alpha}^{(i)}, i = 1,2,3$ are expected to be convex and those on $\widehat{S}^{(i)}, i = 1,2,3$ to be multi-well, nonconvex.

The balances of linear and angular momentum are given by
\begin{equation}\label{eqn:lin_ang_bal}
\begin{split}
\rho \dot{v} & = div \, T + b = \rho \ddot{u}\\
0 & = div \, \Lambda - X \inprod{2} T + K
\end{split}
\end{equation}
where $\rho$ is the mass density, $v$ is the material velocity vector, $T$ is the stress, $\Lambda$ is the couple stress, and $b,K$ are the body force and body-couple densities per unit volume, respectively. As usual in solid mechanics, we assume balance of mass is satisfied once the deformation map at any instant is determined by evaluating the density field on the deforming body from the formula $\rho = \frac{\rho_0}{det(I + \nabla u)}$, where $\rho_0$ is the density field on the reference configuration.

The mechanical power supplied to the body is defined as \cite{mindlin1962effects}
\begin{equation*}
\begin{split}
{\sf P} & := \int_B b \inprod{1} v \,dv + \int_{\partial B} (T n) \inprod{1} v \, da + \int_{\partial B} (\Lambda n ) \inprod{1} \omega \, da + \int_B K \inprod{1} \omega \, dv\\
& = \int_B \rho v \inprod{1} v \, dv + \int_B \left[ T \inprod{2} D + \Lambda \inprod{2} M \right] \, dv,
\end{split}
\end{equation*}
using the balances of linear and angular momentum, where $n$ is the outward unit normal to the boundary of the body, $\omega := \half curl v = -\half X \inprod{2} \Omega$ is the rotation vector where $\Omega := \half \left( \nabla v - (\nabla v)^T \right)$ is the rotation-rate tensor, $D := \half \left( \nabla v + (\nabla v)^T \right) $ is the strain-rate tensor, and $M := \nabla \omega$. Denoting
\begin{equation*}
{\sf F} = \int_B \psi \, dv; \qquad \qquad  {\sf K} = \int_B \half \rho v \inprod{1} v \, dv
\end{equation*}
the \emph{mechanical dissipation}, $\sf D$, or the difference between the power supplied to the body and that stored in it, is given by
\begin{equation}\label{eqn:diss}
{\sf D} := {\sf P} - \dot{\overline{\sf K + \sf F}} = \int_B \left( T \inprod{2} D + \Lambda \inprod{2} M - \dot{\psi} \right) \, dv.
\end{equation}
In the following, we deduce guidelines for constitutive specification in our model that ensure that the mechanical dissipation vanishes in the absence of eigenwall and defect field evolution in any process and is positive otherwise, a minimal necessary condition for the mathematical model to be well-posed.

To facilitate the derivation of the thermodynamic driving forces for the various defect density and eigenwall fields, we will need the following auxiliary fields $P^{(i)}, i \in \{2,3\}$ defined by the solutions of the following Poisson equations:
\begin{equation}\label{eqn:P}
\begin{rcases}
div \, \nabla P^{(i)}&  =  \partial_{\widehat{H}^{(i)}} \psi  \ \qquad x \in B \\
\nabla P^{(i)}\,n & = 0 \qquad \qquad x \in \partial B
\end{rcases} \qquad  i \in \{2,3\},
\end{equation}
which requires that the free-energy density function should satisfy the constraint
\[
\int_B \partial_{\widehat{H}^{(i)}} \psi \, dv = 0, \qquad i \in \{2,3\}.
\]
(This is formally easily arranged by taking any arbitrary $\tilde{\psi}$ with the dependencies of \eqref{eqn:psi}$_3$, and defining $\psi = \tilde{\psi} - \sum_{i = 2}^3 \left( |\Omega|^{-1} \int_\Omega  \partial_{\widehat{H}^{(i)}} \tilde{\psi} \, dv \right) \inprod{i}\widehat{H}^{(i)}$, but its physical and rigorous mathematical implications need to be understood).

Defining $R^{(i)} := \partial_{\chi^{(i)}} \psi$, the fields $W_{R^{(i)}}$ satisfying
\begin{equation}\label{eqn:WR}
    \begin{rcases}
 curl \, curl \, W_{R^{(i)}} =- div \, \nabla \, W_{R^{(i)}} & = curl \, R^{(i)} \qquad x \in B \\
    div \,  W_{R^{(i)}} & = 0 \quad  \qquad \qquad x \in B \\
    W_{R^{(i)}} \times n &= 0 \quad\qquad \qquad x \in \partial B 
    \end{rcases}
    \qquad i \in \{2,3 \}
\end{equation}
(that exist by a unique Stokes-Helmholtz resolution of $R^{(i)}$), will aso be required in the sequel for deriving the thermodynamic driving forces.

A long computation involving \eqref{eqn:psi}$_3$ and the kinematics of the model defined in Sec. \ref{sec:kin} reveals that the mechanical dissipation may be expressed in the suggestive form
\begin{align}\label{eqn:diss_exp}
{\sf D} = & \int_B \left[ T - \partial_{\,\nabla u} \psi + div\, \partial_{\, \nabla^2 u} \psi - div\, div\, \partial_{\, \nabla^3 u} \psi\right]^{(s)} \inprod{2} D \, dv\\
& + \bigintss_B \left[ -\half X \inprod{1} \Lambda^{dev} - \partial_{\, \nabla^2 u} \psi + div\, \partial_{\, \nabla^3 u} \psi \right]^{(a)} \inprod{3} \nabla \Omega \, dv\\
& + \int_{\partial B} \left[ - \partial_{\, \nabla^2 u} \psi \,n + (div \,\partial_{\, \nabla^3 u} \psi)\, n    \right]^{(s)} \inprod{2} D \, da \ +  \ \int_{\partial B} \left[ - \partial_{\, \nabla^3 u} \psi \, n \right] \inprod{3} \nabla^2 v \, da\\
& + \bigintss_B \sum_{i = 1}^3  \left[ X \left( \left( - \partial_{\widehat{S}^{(i)}} \psi + curl\, \partial_{\widehat{\alpha}^{(i)}} \psi   \right)^T \inprod{i} \widehat{\alpha}^{(i)} \right) \right]  \inprod{1} V^{\parallel (i)} \, dv\\
& + \bigintss_B \sum_{i = 1}^3 \left[ \left( div \, \partial_{\widehat{S}^{(i)}} \psi \right) \inprod{i} \widehat{S}^{(i)} \right] \inprod{1} V^{\perp (i)} \, dv\\
& + \bigintss_{\partial B} \sum_{i = 1}^3 \left[ X \left( \left( \partial_{\widehat{\alpha}^{(i)}} \psi \times n   \right)^T \inprod{i} \widehat{\alpha}^{(i)} \right) \right] \inprod{1} V^{\parallel (i)} \, da\\
& + \bigintss_{\partial B} \sum_{i = 1}^3 \left[ - \left( \partial_{\widehat{S}^{(i)}} \psi \, n \right) \inprod{i} \widehat{S}^{(i)} \right] \inprod{1} V^{\perp (i)} \, da\\
& + \bigintss_B \sum_{i=2}^3 \left[ X \left( \left( \nabla P^{(i)} \right)^T \inprod{i} \widehat{\alpha}^{(i)} \right) \right] \inprod{1} V^{\parallel (i)} \, dv \\
& + \bigintss_B \sum_{i=2}^3 \left[  \left( - \partial_{\widehat{H}^{(i)}} \psi \right) \inprod{i} S^{(i)} \right] \inprod{1} V^{\perp (i)} \, dv\\
& + \bigintss_B \sum_{i=2}^3 \left[ X \left( \bigg( curl \, W_{R^{(i)}} \bigg)^T \inprod{i} \widehat{\alpha}^{(i)} \right) \right] \inprod{1} V^{\parallel (i)} \, dv.
\end{align}
Thus, a set of constitutive equations, driving forces for dissipative mechanisms (denoted below by the symbol $\leadsto$), and some boundary conditions for the model are
\begin{equation}\label{eqn:stress}
 T^{(s)} = \left[ \partial_{\,\nabla u} \psi - div\, \partial_{\, \nabla^2 u} \psi + div\, div\, \partial_{\, \nabla^3 u} \psi\right]^{(s)}
\end{equation}
\begin{equation}\label{eqn:couple_stress}
\Lambda^{dev} =  - X \inprod{2} \left[ \partial_{\, \nabla^2 u} \psi - div\, \partial_{\, \nabla^3 u} \psi \right]^{(a)}
\end{equation}
\begin{equation}\label{eqn:hobc1}
 \left. \left[ - \partial_{\, \nabla^2 u} \psi \,n + (div \,\partial_{\, \nabla^3 u} \psi)\, n    \right]^{(s)} \right\vert_{\partial B} = 0
\end{equation}
\begin{equation}\label{eqn:hobc2}
\left. \left( \partial_{\, \nabla^3 u} \psi \right) \, n \right\vert_{\partial B} = 0
\end{equation}
\begin{equation}\label{eqn:Vbulk}
\begin{split}
& \begin{rcases}
V^{\parallel (i)} &  \leadsto  X \left( \left( - \partial_{\widehat{S}^{(i)}} \psi + curl\, \partial_{\widehat{\alpha}^{(i)}} \psi   \right)^T \inprod{i} \widehat{\alpha}^{(i)} \right)\\
V^{\perp (i)} & \leadsto \left( div \, \partial_{\widehat{S}^{(i)}} \psi \right) \inprod{i} \widehat{S}^{(i)}\\
\end{rcases}, \qquad i = 1 \\
& \begin{rcases}
V^{\parallel (i)} & \leadsto  X \left( \left( - \partial_{\widehat{S}^{(i)}} \psi + curl\, \partial_{\widehat{\alpha}^{(i)}} \psi   + \nabla P^{(i)} + curl \, W_{R^{(i)}} \right)^T \inprod{i} \widehat{\alpha}^{(i)} \right)\\
V^{\perp (i)} & \leadsto \left( div \, \partial_{\widehat{S}^{(i)}} \psi - \partial_{\widehat{H}^{(i)}} \psi \right) \inprod{i} \widehat{S}^{(i)}
\end{rcases}, \qquad i = 2,3\\
\end{split}
\end{equation}
\begin{equation}\label{eqn:Vbc}
\begin{rcases}
\left. V^{\parallel (i)} \right\vert_{\partial B} & \leadsto X \left( \left( \partial_{\widehat{\alpha}^{(i)}} \psi \times n   \right)^T \inprod{i} \widehat{\alpha}^{(i)} \right)\\
\left. V^{\perp (i)} \right\vert_{\partial B} & \leadsto - \left( \partial_{\widehat{S}^{(i)}} \psi \, n \right) \inprod{i} \widehat{S}^{(i)}
\end{rcases}, \qquad i = 1,2,3
\end{equation}
(it can be checked that the rhs of \eqref{eqn:couple_stress} is deviatoric).
Equations \eqref{eqn:stress}-\eqref{eqn:hobc2} along with the constitutive choices for the defect and eigenwall velocities to be in the direction of their respective driving forces, mediated by a positive, mobility/drag scalar required on dimensional grounds, ensures non-negative dissipation. Of course, other choices consistent with positive dissipation are possible as well. The boundary conditions \eqref{eqn:hobc1}-\eqref{eqn:hobc2} are not the most general, but a compromise between including higher order stress tensors with dubious physical meaning beyond couple stresses and simplicity in an already involved higher order theory of defects.

It is clear from \eqref{eqn:stress}-\eqref{eqn:couple_stress} and \eqref{eqn:lin_ang_bal} that the governing equations lead to sixth-order pde in the displacement field $u$ (see Sec. \ref{sec:example} below).
\begin{remark}
A minimal set of field variables to be evolved in the model are $(u, \widehat{S}^{(i)}, i = 1,2,3)$ governed by  \eqref{eqn:lin_ang_bal} and \eqref{eqn:S_evol_spec}, with $\widehat{H}^{(1)}, \widehat{H}^{(2)}$ determined from \eqref{eqn:SH_Shat}, $W_{R^{(i)}}, i = 2,3$ determined from \eqref{eqn:WR}, and $\widehat{\alpha}^{(i)}, i = 1,2,3$ determined from \eqref{eqn:defalpha}.
\end{remark}
\begin{remark}
The composite eigenwall fields are coupled to each other through \eqref{eqn:S_evol_spec} and through the displacement field, appearing in the driving forces for the defect and eigenwall velocity fields, governed by \eqref{eqn:lin_ang_bal}. The results of Sec. \ref{sec:motiv} shows how the presence of a higher order defect (characterized by $\Gamma \neq 0$ in a non-simply connected domain) induces a lower order defect \emph{(}$\llbracket y \rrbracket \neq 0$\emph{)} that, in general, induces stress in the body \emph{(}Remark \ref{rem:defect_stress}\emph{)}.
\end{remark}
\begin{remark}
A theory of only surfaces of inflection and singularities arises by assuming $\widehat{S}^{(1)} = 0$ and $\widehat{S}^{(2)} = 0$. A theory of only dislocations arises by setting $\widehat{S}^{(3)} = 0$ and $\widehat{S}^{(2)} = 0$ along with $V^{\perp (2)} = 0$. A theory of only g.disclinations arises by setting  $\widehat{S}^{(1)} = 0$ and $\widehat{S}^{(3)} = 0$ along with $V^{\perp (3)} = 0$. Pair-wise coupled defect theories \emph{(}dislocations $+$ g.disclinations, dislocations $+$ branch/inflection defects, g.disclinations $+$ branch/inflection defects\emph{)} can be obtained by similar means.
\end{remark}

\section{Example: a model of branch-point defects in an elastic body}\label{sec:example}
We specialize the general formalism to a specific case by making the simplest possible choice for the free energy density \eqref{eqn:psi}:
\begin{equation}\label{eqn:spec_psi}
\psi = \half (\nabla u) C (\nabla u) + \half c_2 \left| \nabla^2 u \right|^2 + \half c_3 \left| \nabla^3 u - S \right|^2 + d_3 f ( l^2 |S|) + \half \epsilon_3 \left| curl S \right|^2,
\end{equation}
with the ansatz that $\widehat{S}^{(1)} = 0$, $\widehat{S}^{(2)} = 0$, so that $\widehat{S}^{(3)} = S^{(3)} =:S$. Here, $C$ is the $4^{th}$-order tensor of elastic moduli with major and minor symmetries, $c_2, c_3$ are non-negative scalars (in place of sixth and eighth order tensors!), $d_3$ is a positive scalar (that could also be a positive scalar-valued function of $|curlS|$), and $l, \eps_3$ are positive scalars. The physical dimensions of $c_2, c_3,d_3, l, \eps_3$ are $stress.(length)^2$, $stress.(length)^4$, $stress$, $length$, and $stress.(length)^6$, respectively. Since the equilibria we envisage are of nominally elastic bodies that show non-trivial shapes under no applied loads, $f$ is generally expected to be a multi-well nonconvex function with the bottom of one well at the argument $0$.

Thus we are looking for the mechanics of surfaces of inflection and branch line defects in bodies with an evolving stress-free reference characterized by the choices $\widehat{S}^{(1)} = 0$, $\widehat{S}^{(2)} = 0$, $\widehat{S}^{(3)} = S$, and since it is impossible to construct a displacement field of a 3-d body with vanishing strain, i.e., $\left(\nabla u)^{(s)} = 0\right)$, whose third gradient is non-vanishing, the energy/stress-free reference for our body is never immersible in three-dimensional Euclidean space whenever $S \neq 0$, {i.e.} the stress-free state  is necessarily incompatible or non-realizable.

The balances of linear and angular momentum \eqref{eqn:lin_ang_bal} are solved by taking a $curl$ of \eqref{eqn:lin_ang_bal}$_2$ to obtain
\begin{equation*}
    div \, T^{(a)} = \half\, curl \left( div \, \Lambda^{dev} \right) + \half curl \, K,
\end{equation*}
that on substitution in \eqref{eqn:lin_ang_bal}$_1$ leads to
\begin{equation}\label{eqn:red_lin_mom}
    \rho \ddot{u} = div\, T^{(s)} + \half \, curl \left( div \, \Lambda^{dev} \right) + \half curl \, K + b.
\end{equation}
Constitutive equations \eqref{eqn:stress}-\eqref{eqn:couple_stress} are used to solve for a displacement field from \eqref{eqn:red_lin_mom} (when the defect fields are assumed given), thus satisfying \eqref{eqn:lin_ang_bal}$_1$, and \eqref{eqn:lin_ang_bal}$_2$ is then satisfied, in terms of this displacement field, by simply evaluating $T^a$ from the equation
\begin{equation}\label{eqn:Ta}
    X \inprod{2} T^{(a)} - \frac{1}{3} \nabla (tr\Lambda) = div \, \lambda^{dev} + K,
\end{equation}
making the assumption that the constitutively undetermined $tr\Lambda = 0$, without loss of generality.

For the constitutive choice \eqref{eqn:spec_psi}
\begin{align}
    \Lambda^{dev} & = -c_2 \, X \inprod{2} \left(\nabla^2 u \right)^{(a)} + c_3\, X\inprod{2} \left(div (\nabla^3 u) \right)^{(a)} - X \inprod{2} (div \, S)^{(a)}; \label{eqn:lambda_dev}\\
     \Lambda^{dev}_{il}  & =  e_{ijk} \left( -c_2 \,u_{[j,k]l} + c_3 \, u_{[j,k]lmm} - c_3 \,S_{[jk]lm,m}  \right); \notag\\
     \half \left( curl \, \left(div \, \Lambda^{dev}\right) \right)_i & = - c_2 \, u_{[i,m]llm} + c_3 \, u_{[i,m]llppm} - c_3 \, S_{[im]lp,plm} \notag
\end{align}
and
\begin{align}
        T^{(s)} & = C \nabla u - c_2 \left( div \, \nabla^2 u \right)^{(s)} + c_3 \left( div \, div \, \nabla^3 u \right)^{(s)} - c_3 \left( div \, div \, S \right)^{(s)}; \label{eqn:Ts}\\
        T^{(s)}_{im} & = C_{imkl}\, u_{k,l} - c_2 \, u_{(i,m)ll} + c_3 \, u_{(i,m)lppl} - c_3\, S_{(im)lp,pl}; \notag\\
        \left( div \, T^{(s)} \right)_i & = C_{imkl}\, u_{k,lm} - c_2 \, u_{(i,m)llm} + c_3 \, u_{(i,m)lpplm} - c_3\, S_{(im)lp,plm} \notag
\end{align}
so that the governing equation for the displacement field (\eqref{eqn:red_lin_mom}) may be written as
\begin{equation}\label{eqn:displ_gov}
\rho \ddot{u} = c_3 \, \Delta^3 u - c_2 \, \Delta^2 u + div \, ( C \nabla u) - c_3\, div \, div \, div \, S + \half curl \, K + b,
\end{equation}
where $\Delta^3$ ($\Delta^3(\cdot) = (\cdot)_{,iijjkk}$) and  $\Delta^2$ ($\Delta^2(\cdot) = (\cdot)_{,iijj}$) are the \emph{triharmonic}  and the \emph{biharmonic}  operators, respectively.

To develop the evolution equation for the field $S$ we assume $V^{\perp(3)} = 0$ for simplicity. Since $\psi$ in \eqref{eqn:spec_psi} does not depend on $H^{(3)}$, we have $P^{(3)} = 0$ in \eqref{eqn:Vbulk}$_3$. The governing equation for the evolution of $S$ therefore is given by
\begin{equation}\label{eqn:S_gov}
    \dot{S} = \frac{1}{B} \, curl \, S \times \left( X \left(   \left( c_3\, (\nabla^3 u - S) - d_3 \,l^2 f' \!\left(l^2|S| \right) \frac{S}{|S|} - \epsilon_3 \,curl \, curl \, S \right)^T \inprod{3} curl \, S \right)   \right),
\end{equation}
where $B$ is a drag coefficient with physical dimensions of $stress. (length)^{-2}. time$. 

\begin{remark}
Spatial derivatives of the 3-eigenwall field serve as a source term in \eqref{eqn:displ_gov}; for instance, if $S(x) = g(\nu \cdot x) \, b \otimes \nu \otimes \nu \otimes \nu$, where $\nu$ is the unit normal to a planar surface, $g$ is a scalar-valued function of the spatial coordinate along $\nu$ given by $\zeta = \nu \cdot x$ \emph{(}say a Gaussian centered at $\zeta = 0$\emph{)}, and $b$ is a constant vector, this forcing is of the form $\frac{d^3 g}{d\zeta^3} b$.

Equation \eqref{eqn:S_gov} implies that there is no evolution of the eigenwall field at locations where $curl\, S = 0$, regardless of the energetic driving force there. For example, the field $S(x) = g(\nu \cdot x) \, b \otimes \nu \otimes \nu \otimes \nu$ has no `longitudinal' variation and does not evolve according to \eqref{eqn:S_gov}. However, $S(x) = g(t \cdot x) g(\nu \cdot x) \, b \otimes \nu \otimes \nu \otimes \nu$, where $t$ is orthogonal to $\nu$ does evolve. Physically, the eigenwall field is `dragged' by the evolution of its core.
\end{remark}
\begin{remark}
The governing equation \eqref{eqn:displ_gov} implies that, when the elastic modulus $C$ is homogeneous and isotropic, given by $C_{ijkl} = \lambda u_{k,k} \delta_{ij} + \mu (\delta_{ik} \delta_{jl} + \delta_{il}\delta_{jk})$, plane waves of $curl \, u$ and $div \, u$ are dispersive in nature, with propagation possible in any direction in space. The dilatational waves (i.e., waves of $div \, u$) with wave number $|k|$ and direction $\frac{k}{|k|}$ propagate with velocity
\[
c_d := \pm \sqrt{\frac{c_3 |k|^4 + c_2 |k|^2 + (\lambda + 2\mu)}{\rho}}
\]
while the equivoluminal waves or `shear waves' (i.e., vectorial waves of $curl \, u$) propagate with velocity 
\[
c_s := \pm \sqrt{\frac{c_3 |k|^4 + c_2 |k|^2 + \mu}{\rho}}.
\]
Continuous dependence w.r.t initial data of the Cauchy problem for the evolution of displacement requires $c_3 \geq 0$. When $c_3 = 0$, $c_2$ must be non-negative, with the requirement that $ \mu \geq 0$ and $\lambda + 2 \mu \geq 0$ if $c_2 = 0$. Within these parameter regimes, linear instabilities can arise for wavenumber and parameter combinations resulting in $c_d$ or $c_s$ taking complex values.
\end{remark}
\subsection{Uniqueness of the displacement field and boundary conditions}\label{sec:bc}
Our model encompasses a model of third-order elasticity in the absence of dissipative defect evolution, and involves the thermodynamically motivated higher-order boundary conditions \eqref{eqn:hobc1}-\eqref{eqn:hobc2}. Here, we use a uniqueness argument (in a putative smooth class of solutions) to deduce a full set of boundary conditions for the problem \eqref{eqn:displ_gov} when the $S$ field is assumed specified. We abstract the results of the exercise in this special case related to the `quadratic' energy \eqref{eqn:spec_psi} to identify a likely set of sufficiently general boundary conditions for the determination of the displacement field for processes consistent with the general constitutive statement \eqref{eqn:psi}.

Consider two solutions $u^{(1)}$ and $u^{(2)}$ of \eqref{eqn:displ_gov} corresponding to identical $S, K, b$ fields. Denote the \emph{difference displacement} as $u := u^{(1 )}- u^{(2)}$ and its velocity $v = \dot{u}$. Then  $u$ satisfies
\begin{equation*}
\rho \ddot{u} = c_3 \, \Delta^3 u - c_2 \, \Delta^2 u + div \, ( C \nabla u),
\end{equation*}
and taking the inner-product of the difference velocity with the equation and integrating in space, we have
\[
\half \frac{d}{dt} \int_B \rho \, v_i v_i \, dv  = \int_B C_{imkl} \, u_{k,lm} v_i \, dv - \int_B c_2 u_{i,mmll} \,  v_i \,dv + \int_B c_3 u_{i,mmllpp} \, v_i \, dv,
\]
which implies
\begin{align}\label{eqn:energy_equality}
   & \half \frac{d}{dt} \int_B \rho \, v_i v_i \, dv + \int_B C_{imkl} \, u_{k,l}  v_{i,m} \, dv +  \int_B c_2 u_{i,ml} \,  v_{iml} \,dv + \int_B c_3 u_{i,mlp} \, v_{i,mlp} \, dv \notag\\
    & = \quad \! \int_{\partial B} \big( C_{ilkm}\, u_{k,m} - c_2 \, u_{i,mml} + c_3 \, u_{i,mmppl} \big) v_i \,n_l \, da  \notag\\
    & \quad +  \int_{\partial B} \big( c_2 \, u_{i,lm} - c_3 u_{i,lppm} \big) v_{i,l} \, n_m \, da \notag\\
    & \quad + \int_{\partial B}  \big( c_3 \, u_{i,plm} \big) \, v_{i,lp} \, n_m \, da.
\end{align}
Let us now assume that both $u^{(1)}$ and $u^{(2)}$ satisfy \eqref{eqn:hobc1}-\eqref{eqn:hobc2} consistent with \eqref{eqn:spec_psi}. Then the last line of \eqref{eqn:energy_equality} vanishes due to the boundary condition \eqref{eqn:hobc2} and the line before that due to \eqref{eqn:hobc1}.

Let the stress field arising from $(u^{(i)}, S)$, $i = 1 \, \mbox \, 2$, be $T^{(i)} = T^{(s)(i)} + T^{(a)(i)}$, in accord with \eqref{eqn:Ta}, \eqref{eqn:lambda_dev}, and \eqref{eqn:Ts}. Then the third line from the bottom of \eqref{eqn:energy_equality} may be interpreted as
\begin{equation*}
     \int_B \Big( T^{(1)}_{il} - T^{(2)}_{il} \Big) v_i \, n_l \, da
\end{equation*}
and if we now additionally require that solutions satisfy specified tractions and velocities (or displacements) on mutually complementary parts of the boundary of the body for all times, then this term vanishes.

Consequently, we are left with
\begin{equation*}
    \frac{d}{dt} \left( \half \int_B \rho \, v_i v_i \, dv + \half \int_B C_{imkl} \, u_{k,l}  u_{i,m} \, dv + \half \int_B c_2 u_{i,ml} \,  u_{iml} \,dv + \half \int_B c_3 u_{i,mlp} \, u_{i,mlp}\, dv \right) = 0
\end{equation*}
and if $u^{(1)}$ and $u^{(2)}$ both satisfy specified initial conditions on the displacement and velocity fields, then the bracketed quantity, an integral of sums of squares (in fact, the potential and kinetic energies of the body subjected to the difference displacement) vanishes at all times. This proves that the difference velocity vanishes point-wise, and the initial condition on the difference displacement implies that $u^{(1)} = u^{(2)}$ for all $(x,t)$. Obviously, the dynamic problem allows the prediction of unique rigid motions. In statics, i.e., when the inertia term is absent, one takes the inner product of the governing equation for the difference displacement with the difference displacement, and obtains, for the same boundary conditions (except only the displacement can now be specified on the part of the boundary complementary to where tractions are specified),
\begin{equation*}
   \int_B C_{imkl} \, u_{k,l}  u_{i,m} \, dv + \int_B c_2 u_{i,ml} \,  u_{iml} \,dv + \int_B c_3 u_{i,mlp} \, u_{i,mlp} \, dv  = 0.
\end{equation*}
All integrands are non-negative implying that the strain, or the symmetrized displacement gradient, vanishes (recall the minor symmetries of $C$) which, by compatibility, further implies that the displacement field is unique if a displacement boundary condition is specified and otherwise it is unique up to an infinitesimally rigid deformation.

Thus, the higher order boundary conditions \eqref{eqn:hobc1}-\eqref{eqn:hobc2}, along with classical displacement and traction boundary conditions may be expected to define a well-set problem (for the displacement field) in the case of the general constitutive equation \eqref{eqn:psi}. Of course, the traction now involves a stress tensor that has an antisymmetric part, and is constitutively dependent on higher order displacement gradients.
\subsection{A `plate' idealization}
Let the reference $B$ be a plate of thickness $2t$, i.e., $B = \{(x_1,x_2,x_3)| (x_1,x_2,0) \in B_2, x_3 \in [-t, +t]\}$, where $B_2$ is a flat 2-dimensional simply connected domain. Defining the through-the-thickness average of a function as
\[
\overline{f}(x_1,x_2) := \frac{1}{2t} \int_{-t}^{+t} f(x_1,x_2,x_3)\, dx_3
\]
and the notation
\[
\left[ f \right]^{+t}_{-t} (x_1, x_2) := f(x_1, x_2, +t) - f(x_1, x_2, -t),
\]
we now seek the governing equations for $\overline{u}$ and $\overline{S}$, under the ansatz that $\overline{S} = S$ and $\overline{\rho} = \rho$, i.e., $S$ and $\rho$ do not vary through the thickness of the plate, and $K = b = 0$. It is also assumed that a component of $S$ vanishes if any of its last three indices takes the value $3$. We use the notation that all lowercase Greek indices vary from $1$ to $2$ while lowercase Latin indices span from $1$ to $3$.

While not essential, the assumptions $l = 2t$, $c_2 = E t^2$ and $c_3 = E t^4$, where $E$ is the Young's modulus of the material can be made to draw an analogy with classical plate theory (the curvature-related elastic energy term in the thickness-integrated expression of \eqref{eqn:spec_psi} would then be proportional to $t^3$). For $0 < t  \ll 1$, whenever $S \neq 0$, there is energy and stress in the body, possibly small, with the corresponding thickness-integrated `elastic' energy of the plate (arising from the first three terms in \eqref{eqn:spec_psi}), alternatively the `plate elastic energy', scales as $\sim t^5$, assuming energy is minimized, there are no external forcing or constraints, and $\epsilon_3 >0$ to rule out any possibility of a singular energy. Our governing equations \eqref{eqn:displ_gov} or \eqref{eqn:u_avg} do not {\em require} that energy be minimized, so that scaling of the thickness-integrated elastic energy w.r.t $t$ as $t \to 0$ in the model can well contain lower order bending $\left(O(t^3)\right)$, and even stretching $\left(O(t)\right)$, contributions.

Applying the averaging operator to \eqref{eqn:displ_gov} and noting that
\begin{equation*}
    \begin{split}
        u_{i,llppmm} &= u_{i,\alpha \alpha \beta \beta \gamma \gamma} + 3 u_{i,\beta \beta \gamma \gamma 33} + 3 u_{i, \gamma \gamma 3333} + u_{i,333333}\\
        u_{i,llpp} &= u_{i,\alpha \alpha \beta \beta} + 2 u_{i,\beta \beta 33} + u_{i,3333}\\
        C_{ijkl}u_{k,lj} &= C_{i\beta k \alpha} u_{k,\alpha \beta} + C_{i \beta k 3} u_{k, \beta 3} + C_{i3k \alpha} u_{k,\alpha 3} + C_{i3 k \alpha} u_{k,33}\\
        S_{ijkl,jkl} &= S_{i \alpha \beta \gamma, \alpha \beta \gamma} + \left( S_{i \alpha \beta 3} + S_{i3 \alpha \beta} + S_{i \alpha 3 \beta} \right)_{,\alpha \beta 3} + \left( S_{i333 \gamma} + S_{i3 \gamma 3} + S_{i \gamma 33} \right)_{,33 \gamma} + S_{i333,333},
    \end{split}
\end{equation*}
we obtain
\begin{equation}\label{eqn:u_avg}
\begin{split}
    \rho \ddot{\overline{u_i}} &=  c_3 \, \overline{u_i}_{,\alpha \alpha \beta \beta \gamma \gamma} - c_2 \, \overline{u_i}_{,\alpha \alpha \beta \beta} + C_{i \beta k \alpha}\overline{u_k}_{,\alpha \beta} - c_3 \overline{S_{i \alpha \beta \gamma}}_{,\alpha \beta \gamma}\\
    & \quad + \big[ 3 c_3 u_{i, \gamma \gamma 333} + c_3 u_{i,33333} - 2 c_2 u_{i,\beta \beta 3} - c_2 u_{i,333} + C_{i \beta k 3} u_{k,\beta} + C_{i3k \beta} u_{k,\beta} + C_{i3 k 3} u_{k,3} \big]^{+t}_{-t}.
\end{split}
\end{equation}
Similarly,
\begin{equation}\label{eqn:S_avg}
\begin{split}
   B \, \dot{\overline{S_{i \pi \sigma \lambda}}} & = e_{3 \mu \rho} \,\overline{S_{i \pi \sigma \rho}}_{,\mu} \, e_{\lambda 3 \chi}\left(e_{\chi \xi 3} \Bigg\{ c_3 \Big( \overline{u_w}_{,\alpha \beta \xi} - \overline{S_{w \alpha \beta \xi}} \Big) \right.\\
    & \qquad \qquad \qquad \qquad \qquad \quad- \epsilon_3 \, e_{\xi \nu 3} \, e_{3 \gamma \phi} \overline{S_{w \alpha \beta \phi}}_{,\gamma \nu} - d_3 \, l^2 f'(l^2 |\overline{S}|) \frac{\overline{S_{w \alpha \beta \xi}}}{\left|\overline{S} \right|}  \Bigg\}  e_{3 \epsilon \zeta}\, \overline{S_{w \alpha \beta \zeta}}_{,\epsilon} \Bigg)\\
    & \quad \ + e_{3 \mu \rho} \,\overline{S_{i \pi \sigma \rho}}_{,\mu} \, e_{\lambda 3 \chi} \, e_{\chi \xi 3} \, e_{3 \epsilon \zeta} \left( \overline{S_{w \alpha 3 \zeta}}_{,\epsilon} \left[ u_{w, \alpha \xi} \right]^{+h}_{-h} +  \overline{S_{w 3 \beta \zeta}}_{,\epsilon} \left[ u_{w, \beta \xi} \right]^{+h}_{-h} + \overline{S_{w 3 3 \zeta}}_{,\epsilon} \left[ u_{w, 3 \xi} \right]^{+t}_{-t} \right).
\end{split}
\end{equation}
In equation \eqref{eqn:u_avg}, the terms beyond the first line represent forcings in the transverse direction to the plate and need to be specified (it would be physically legitimate to assume many of these terms to vanish); the third line of \eqref{eqn:S_avg} has similar meaning and needs specification.

The functions $\overline{u}, \overline{S}$ represent the fundamental fields of the plate theory, governed by \eqref{eqn:u_avg}-\eqref{eqn:S_avg}. Evaluating $\overline{T^{(a)}}$ from \eqref{eqn:Ta} in terms of $(\overline{u}, \overline{S})$ solving \eqref{eqn:u_avg}-\eqref{eqn:S_avg} and $\overline{K}$ would imply the satisfaction of balance of angular momentum (i.e., moment balance) in the through-the-thickness averaged sense.
\begin{remark}
We note that non-evolving and non-vanishing $\widehat{S}^{(1)}, \widehat{S}^{(2)}$ `target' composite eigenwall fields can be included in the considerations of this Section (Sec. \ref{sec:example}), with only slight increase of tedium in bookkeeping.

Within the context of energy minimization and for $t >0$, if $curl \, \left(curl\, \left(\widehat{S}^{(1)(s)}\right)\right)^T = 0$, i.e. $\widehat{S}^{(1)(s)}$ satisfies the St.-Venant compatibility condition, then an infinitesimal isometry exists (the reference configuration is assumed to be simply-connected) and the plate elastic energy scales as $\sim t^3$ or of smaller magnitude; if $\widehat{S}^{(1)(s)}$ is not compatible, then the energy has to scale as $\sim t$. We note that when $\widehat{S}^{(1)(s)}$ is compatible, unless $\widehat{S}^{(2)} = \nabla^2 v$, where $v$ is s.t. $(\nabla v)^{(s)} = \widehat{S}^{(1)(s)}$ so that $\nabla^2 v$ is unique, the plate elastic energy is going to scale as $\sim t^3$. The requirement $\widehat{S}^{(2)} = \nabla^2 v$ is non-generic for a freely-specifiable $\widehat{S}^{(2)}$ field that, however, is satisfied by the choice $\widehat{S}^{(1)} = 0, \widehat{S}^{(2)} = 0$. Thus, in most circumstances the plate energy is expected to scale as $\sim t^3$, if the plate energy is minimized.
\end{remark}

\section{Discussion}

Starting from the work of the brothers Cosserat \cite[as presented in \cite{truesdell1960classical}] {cosserat1909theorie}, through those of Toupin \cite{toupin1964theories}, Green and Rivlin \cite{green_rivlin64}, Mindlin \cite{mindlin1962effects,mindlin1964micro},  on to that of Fleck and Hutchinson \cite{fleck1994strain,fleck2001reformulation,hutchinson2012generalizing} and of Gurtin \cite{gurtin2002gradient,gurtin2009thermodynamics}, higher order theories of continuum mechanics have made an appearance off and on and have been noted for their intricacy and elegance, but always, arguably, with the nagging question of the physical justification ({\em in their details }\!\footnote{For example, none of the plasticity-related works in the above, while apparently motivated from modeling plasticity arising from dislocations, recover all of the ingredients of the classical Peach-Koehler force in the driving force for their dislocation-related inelastic deformation mechanisms.}) in view of their added complexity. Our work aims to provides a concrete, tangible, and compelling justification - that the precise treatment of  defects in the deformation and its higher order gradients is the raison d'\^{e}tre for higher order theory in continuum mechanics.

Our work is in the context of non-Euclidean elastic sheets with negative in-plane Gauss curvature. These objects are ubiquitous in nature and they display varied and intricate multi-scale behaviors \cite{eran,audoly2003self,KES07,halftone-gels,GV2013}. Their elastic behavior is significantly different from that of elastic plates or spherical shells \cite{EPL_2016,toby_todo}. In particular, they have ``large" {\em continuous families} of low-energy states obtained from piecewise isometries, with each piece possessing additional ``bending" degrees of freedom. 
Thin  hyperbolic free sheets are thus easily deformed by weak stresses and their morphology is strongly dependent on the dynamics of the growth/swelling processes, material imperfections, or other weak external forces. This naturally motivates the need for tools to describe singularities/defects in these sheets, their interactions and the resulting dynamics. 

Mesoscopic defects in hyperbolic sheets, associated with their ``soft" modes of deformation, include {\em lines of inflection} that terminate at {\em branch points} \cite{GV2012,toby_todo}. These are higher-order defects (termination of jumps in curvature) unlike the more common types of defects, disclinations and dislocations. Irreversible effects in the dynamics of disclinations and dislocations are associated with (macroscopic) plastic behaviors - stress-free large deformations, internal stresses, and microstructure - in solids. A natural question therefore is -- what are the macroscopic manifestations of moving lines/surfaces of inflection and branch points/lines? 

 In this work we have begun to address this question in the context of `small deformations' from a (potentially stressed, when occupied) reference configuration.  A detailed analysis and characterization of the kinematics of branch point defects and the discontinuities in the deformation that they induce is achieved. This analysis, in its essence, is a non-trivial adaptation and extension of the ideas of Weingarten \cite{wein} and Volterra \cite{volt}, from the dawn of elastic defect theory, to a context not restricted within the kinematics of only strain (the symmetrized gradient of the displacement, as well as its nonlinear analog) and its incompatibilities, and shows the natural way forward for deducing the constraints on possible jumps in deformation, i.e. global constraints, for locally compatible higher order deformation gradients, albeit on domains with the simplest non-trivial topology\footnote{It should be noted that the question of conditions for global {\em compatibility} on domains with non-trivial topology is different from the question addressed by Weingarten's theorem and its extensions to higher order kinematics, which deduce constraints on the discontinuous deformations arising from the absence of global compatibility.}. We then develop a thermodynamically consistent theory for the dissipative dynamics of such defects in a nominally elastic solid, allowing for their interaction with dislocation, g.disclination, grain, and phase boundary defects. The constitutive guidance provided by this thermodynamic argument ensures that the model is equipped with an energy (in)equality, a crucial necessary condition for its physical and mathematical well-posedness. The analysis uncovers the non-Newtonian, energetic driving forces on these defects that couple their dynamics and mutual interactions to applied loads and the deformation of the body\footnote{The fact that similar models, for lower-order defect kinematics, can indeed represent the complex nonlinear statics, dynamics, and interaction of defects is demonstrated in \cite{zhang2015single,zhang2016non,zhang2018finite,Arora2018}.}. Evolution of the defect fields subject to such driving forces necessarily reduces the system free-energy by design, within an overall dynamics that accounts for material inertia and is not restricted to its free-energy decreasing with time (depending on the external driving). As an example, we explicitly demonstrate the full set of governing equations for the case of branch point defects in an elastic material and develop a `plate' theory idealization for it. The development of the finite deformation version of the model poses no conceptual or technical barriers\footnote{For the worker proficient in general continuum mechanics.} based on our prior work in g.disclination mechanics \cite{acharya2015continuum}, but this same work makes it clear that the bookkeeping tasks in pushing through the analysis are going to be formidable.
 
 We observe in passing that while we have been interested in developing a theory for branch point/line defects and lines/surfaces of inflection, i.e. a theory for the discontinuities and singularities of the deformation and its gradients up to order three, the analysis makes it clear that the mathematical/continuum mechanical formalism extends to describing the discontinuities and singularities of any finite integer order gradient of the deformation, while including only stresses and couple stresses. As already observed in \cite{acharya2015continuum}, using the Second Law in global form is crucial for this, albeit at the expense of the application of limited (but adequate, as we show in Sec. \ref{sec:bc}) higher-order boundary conditions about which not much is physically known anyway.

As a final comment, we note that a geometric model of growth mechanics, based on Riemannian geometry and including evolution, has been proposed in \cite{yavari2010geometric}. The viewpoint is different from ours and, in particular, the mechanics of incompatibility based on a Riemannian metric cannot describe (without non-trivial extension) the `softer' branch point defects we focus on. We 
expect that one can recast our continuum mechanical kinematic constructs within a differential geometric structure involving the specification of a moving frame, 
{\em and higher-order constructs} based on such a field, thereby making connections with the ``geometric" viewpoint of growth mechanics.

\section*{Acknowledgments}
SCV is supported by the Simons Foundation through awards 524875 and 560103. Portions of this work were carried out when SCV visited the Center for Nonlinear Analysis at Carnegie Mellon University, and their hospitality is gratefully acknowledged.
\bibliographystyle{alpha}
\bibliography{branch_pt}

\end{document}

%% file: temp-definitions.tex
\usepackage{amsmath}
\usepackage{amsbsy}
\usepackage{amssymb}
\usepackage{amscd}
\usepackage{amsfonts}

\mathchardef\mhyphen="2D

\newcommand{\eps}{{\epsilon}}

\newcommand{\beq}{\begin{equation}}
\newcommand{\eeq}{\end{equation}}
\newcommand{\beqs}{\begin{eqnarray}}
\newcommand{\eeqs}{\end{eqnarray}}
\newcommand{\beql}{\begin{equation} \label}
\newcommand{\half}{\frac{1}{2}}

\newtheorem{remark}{Remark}[section]

\newcommand{\Sone}{S^{(1)}}
\newcommand{\Yone}{Y^{(1)}}
\newcommand{\Stwo}{S^{(2)}}
\newcommand{\Ytwo}{Y^{(2)}}
\newcommand{\Sthree}{S^{(3)}}
\newcommand{\Ythree}{Y^{(3)}}
\newcommand{\Shatone}{\widehat{S}^{(1)}}
\newcommand{\Shattwo}{\widehat{S}^{(2)}}
\newcommand{\Shatthree}{\widehat{S}^{(3)}}
\newcommand{\halphaone}{\widehat{\alpha}^{(1)}}
\newcommand{\halphatwo}{\widehat{\alpha}^{(2)}}
\newcommand{\halphathree}{\widehat{\alpha}^{(3)}}
\newcommand{\inprod}[1]{\cdot_{#1}}